\documentclass[epj,nopacs]{svjour}
\usepackage[utf8]{inputenc}
\usepackage{amsmath,amssymb,amsfonts,graphicx,cite,multirow}
\graphicspath{{graphics/}}
\makeatletter
\ifx\input@path\@undefined
\def\input@path{{graphics/}}
\else
\g@addto@macro\input@path{{graphics/}}
\fi
\makeatother

\newcommand{\hw}{\textsf{Herwig~7}}
\usepackage{color}
\usepackage{hyperref}

\preprint{%
MAN/HEP/2016/06\\
CERN-TH-2016-093\\
HERWIG-2016-03\\
IFJPAN-IV-2016-7\\
KA-TP-12-2016\\
MCnet-16-12\\
IPPP/16/34
}

\title{Parton Shower Uncertainties with Herwig 7:\\
Benchmarks at Leading Order}

\author{Johannes Bellm\inst{1} \and Graeme Nail\inst{2,3} \and Simon
  Pl\"atzer\inst{1,2} \and Peter Schichtel\inst{1} \and Andrzej
  Si\'odmok\inst{4,5}}

\institute{ IPPP, Department of Physics, Durham University\and
  Particle Physics Group, School of Physics and Astronomy, University
  of Manchester \and Institute for Theoretical Physics, Karlsruhe Institute of
  Technology \and CERN, TH Department, Geneva\and 
  The Henryk Niewodniczanski Institute of Nuclear Physics in Cracow, Polish Academy of Sciences}

\date{\today}

\abstract{We perform a detailed study of the sources of perturbative uncertainty
  in parton shower predictions within the Herwig 7 event generator. We
  benchmark two rather different parton shower algorithms, based on
  angular-ordered and dipole-type evolution, against each other. We
  deliberately choose leading order plus parton shower as the benchmark
  setting to identify a controllable set of uncertainties. This will enable us
  to reliably assess improvements by higher-order contributions in a follow-up
  work.\PACS{ {xx.yy.zz}{Xx Yy Zz} } }

\begin{document}

\maketitle



\section{Introduction}
\label{sec:intro}

General purpose Monte Carlo (MC) event
generators~\cite{Bahr:2008pv,Bellm:2015jjp,1126-6708-2006-05-026,Sjostrand:2014zea,Sjostrand:2007gs,Gleisberg:2008ta}
are central to both theoretical and experimental collider physics studies.
Recent development of these simulations has seen improvements in various
areas, both within perturbative calculations, through matching to fixed order
\cite{Alioli:2010xd,Platzer:2011bc,Hoeche:2011fd,Larkoski:2013yi,Alwall:2014hca,Hamilton:2013fea,Hoeche:2014aia,Czakon:2015cla,Bellm:2015jjp,Jadach:2015mza},
combining higher jet mutliplicites
\cite{Hoeche:2012yf,Frederix:2012ps,Alioli:2012fc,Lonnblad:2012ng,Platzer:2012bs,Lonnblad:2012ix,Frederix:2015fyz},
as well as the all-order resummation with parton showers
\cite{Platzer:2012np,Nagy:2015hwa,Hoche:2015sya,Fischer:2016ryl} and also  
within the non-perturbative, phenomenological models
\cite{Bierlich:2014xba,Christiansen:2015yqa}. While there are well established
prescriptions on how to quantify the theoretical uncertainty of fixed-order
calculations due to missing higher order contributions
~\cite{Agashe:2014kda,Stevenson:1980du,PhysRevD.28.228,Stewart:2011cf,Cacciari:2011ze,
  David2013266,Bagnaschi:2014wea} \footnote{While being based on scale
  compensation arguments, these methods are, however, not able to predict the
  impact of finite corrections.}, no such general recipe exists for
resummed calculations (see {\it e.g.}~\cite{Bozzi:2005wk,Berger:2010xi}), and
parton shower algorithms in particular. Given the perturbative improvements,
and the expected precision from data-taking at {\it Run II} of the Large Hadron
Collider~\cite{atlas:lumi,cms:lumi}, the task of assigning theoretical
uncertainties to MC event generators is becoming increasingly crucial. This also applies to
validating new approaches against existing data, as well as using predictions to
design future observables and/or collider experiments. Phenomenological
studies, for example, indicate that MC event generators can be used even in primarily data driven
methods to perform powerful analyses once theoretical uncertainties are under
control~\cite{Englert:2011cg,Englert:2011pq}. It is therefore important to
quantify the uncertainties associated with an event generator in a reliable
way.

Uncertainties due to non-perturbative modelling have been addressed
in~\cite{Richardson:2012bn} and~\cite{Seymour:2013qka}, as well as the impact
of the parton shower on reconstructed
observables~\cite{Seymour:2006vv}. Various ambiguities and sources of
uncertainty have been addressed within the context of other multi-purpose
event generators as well; in particular recoil
schemes~\cite{Platzer:2009jq,Hoeche:2014lxa} and parton distribution functions
(PDF)~\cite{Bourilkov:2003kk,Gieseke:2004tc,Buckley:2016caq} have so far been
considered, both for pure showers and in the context of matched or merged
samples, see {\it e.g.}~\cite{Hoche:2012wh}. All of these studies share a
commonality in that they focus on a single source of uncertainty which is
usually connected to the development/improvement studied. Contrary to this,
the authors in~\cite{adam:2008,Krasny:2007cy,Fayette:2008wt,Krasny:2010vd}
describe possible approaches to uncertainty handling for the Drell-Yan
process. An even more systematic approach for how to handle the possible
interplay between theoretical and experimental uncertainties can be found
in~\cite{Cranmer:2013hia}. Further in the direction of a systematic approach,
CMS published a short guide on how to estimate MC
uncertainties~\cite{Bartalini:2005zu} and outlined some issues to
address. Finally, new techniques of propagating uncertainties through the
parton shower by means of an alternate event weight were
proposed~\cite{Stephens:2007ah}.

In the present work, we address uncertainties of parton shower algorithms
within the \hw~event generator \cite{Bahr:2008pv,Bellm:2015jjp}.  \hw~is a
general purpose event generator that computes any observable at
next-to-leading order (NLO) precision in perturbation theory automatically
matched to a parton shower. It includes sophisticated modules for very
different physics aspects ranging from interfaces for physics beyond the
standard model and two independent parton shower
algorithms~\cite{Gieseke:2003rz,Platzer:2009jq}, to a detailed modelling of
multiple particle interactions~\cite{Bahr:2008dy,Bahr:2008wk,Gieseke:2012ft}.

It is our aim to develop a consistent uncertainty evaluation for event
generators, and \hw\ in particular. This work is a first step in this
direction, concerning the parton shower part and will be extended by further
detailed studies in the context of higher order improvements and the interplay
with non-perturbative, phenomenological models and parameter fitting. The
present paper is therefore structured as follows: in Sec.~\ref{sec:context} we
classify all different types of uncertainties and their respective sources. We
then argue as to why we start with a pure leading order (LO) plus parton
shower (PS) study. The sources of uncertainty tested in this study are
described in detail in Sec.~\ref{sec:scales}. Our results are presented in
Sec.~\ref{sec:clean} for $e^+e^-$ and fully inclusive $pp$ production, while
our findings including additional jet radiation are described in
Sec.~\ref{sec:jetty}. The results establish a baseline of a set of
controllable uncertainties, that can then be used to quantify the impact of
higher order corrections to be addressed in upcoming work. Finally, we present
a summary and outlook in Sec.~\ref{sec:outlook}.


\section{Context}
\label{sec:context}

\subsection{Sources of Uncertainty}
\label{subsec:sources}

For any general purpose event generator that is based on both perturbative input and phenomenological
models, there are a number of different sources of uncertainty to be
addressed:
\begin{itemize}
\item {\bf Numerical:} Computational precision and statistical convergence.
  This is clearly a limitation which can be overcome by investing enough
  computing resources and will hence not be addressed further.
\item {\bf Parametric:} Quantities taken from measurements or fits beyond the
  event generator parameters. This includes masses, coupling constants, and PDFs,
  and the impact of these needs to be quantified separately and potentially on
  a process-by-process basis watching out for maximum sensitivity.
\item {\bf Algorithmic:} The actual parton shower algorithm, matching and
  merging prescriptions, and phenomenological models considered. The last are
  not considered here, as we limit ourselves to the simulation available in
  \hw.
\item {\bf Perturbative:} Truncation of expansion series in coupling or
  logarithmic order. The main purpose of this work is to elaborate on
  quantifying these uncertainties in the case of leading order plus
  parton shower simulation, which will be motivated in more detail below.
\item {\bf Phenomenological:} Goodness of fit uncertainties regarding
  parameters in the non-perturbative models. We will argue that a remaining
  spread of predictions obtained by fitting parameters for each of the
  variations of controllable perturbative uncertainties is able to quantify
  the cross talk to non-perturbative models and a genuine model uncertainty.
\end{itemize}

In this study we will address perturbative uncertainties in the parton shower
algorithms as a first piece of the chain of variations to be done. We will use
two different shower algorithms to benchmark the uncertainty prescriptions
against each other and to point out further interesting differences. The
results considered here will serve as further input to identify improvements of
NLO matching and merging, to be addressed in a  separate paper.

Phenomenological uncertainties will be subject to future investigations.
However, we will point out first hints towards their influence by considering
variations of the shower infrared cutoff in a selected number of cases. The
reasoning to this is two-fold: On one hand, we want to stress the
fact that parton level studies should typically be carried out with care, and
their region of validity can by estimated by cutoff variations with large
changes that indicate non-negligible hadronisation corrections. On the other
hand, this fact also indicates how cutoff variations, along with other
variations, may actually point to the possibility of quantifying otherwise
unknown, generic, model uncertainties and the interplay with non-perturbative
corrections.

\subsection{Why Leading Order?}
\label{subsec:leadingorder}

We solely consider LO plus PS simulation in this work. The motivation to do so
is as follows: With fixed-order improvements it is clearly very hard to
disentangle sources of uncertainty stemming from pure parton showering, and
those which have been potentially improved by higher-order corrections. In
order to quantify genuine parton show\-er uncertainties in an improved setting
one would typically need to look at jets beyond those that received
fixed-order hard process input ({\it e.g.} the second jet from a leading order
configuration in a next-to-leading order matched simulation). Not only is this
computationally unnecessary for the sake of studying only parton shower
uncertainties, it also introduces slightly different shower dynamics, the
differences of which, with respect to leading order, would also need to be
quantified carefully. Additionally, it is our aim to show where and how fixed
order input improves the simulation along with the expected reduction in
uncertainty; stated otherwise: To use the NLO matched simulation in order to
identify which of the non-first-principle variations considered in this work
are indeed reliable estimators of theoretical uncertainty in the perturbative
part of event generator predictions.

\subsection{Different Algorithms or Uncertainties?}
\label{subsec:algos}

To quantify to what extent commonly used recipes are a sensible measure of
uncertainties in parton shower algorithms the first step is a clear
distinction of what possible sources exist within a fixed algorithm, and what
differences should actually be attributed to the consideration of distinct
algorithms. Looking at different algorithms, we obtain a strong cross-check on
whether the uncertainties assigned to one algorithm are sensible, provided
we consider algorithms that exhibit similar resummation properties. We will
also show that changes to the algorithms that are naively expected to be
subleading, can cause severe difference in the resummation
properties. Similarly, kinematics parametrisations to convert on-shell partons
to off-shell ones after multiple radiation are known to cause numerically
significant differences \cite{Bengtsson:1986et,Platzer:2009jq,Hoeche:2014lxa}.

Such details, as well as the choice of splitting kernels and evolution
variable should not be considered a source of uncertainty within an algorithm
but are details that fix a distinct algorithm; we therefore call them
algorithmic uncertainties. An uncertainty band based on varying such details
cannot serve as a systematic framework to quantify missing higher-logarithmic
contributions. If differences between algorithms are not covered by variation
of the scales involved, either the estimate of uncertainty {\it or} the
resummation properties of the algorithms should be questioned.

The relevant scales for the study at hand are:
\begin{itemize}
\item the hard scale $\mu_H$ (factorisation and renormalisation scale in the hard process);
\item the veto scale $\mu_Q$ (boundary on the hardness of emissions);
\item the shower scale $\mu_S$ (argument of $\alpha_S$ and PDFs in the parton shower).
\end{itemize}
No a priori prescription can be obtained as to what these scale choices should
optimally be; the first two are usually taken as `a typical scale of the hard
process', while the last one faces more constraints to guarantee resummation
properties and the correct backward evolution \cite{Sjostrand:1985xi} within the
parton shower. Having made a central choice, we vary the scales by fixed
factors to generate subleading terms with coefficients of order one as an
initial guess on higher order corrections and phase-space effects.

At least two parameters in our shower algorithms are typically obtained in the
course of tuning to data, the strong coupling $\alpha_s(M_Z)$, and the shower
cutoff parameter.\footnote{One can argue that the tuning of $\alpha_s(M_Z)$ is
  typically absorbing the CMW correction advertised in \cite{Catani:1990rr}
  which would have to be included otherwise to obtain a satisfactory
  description of data.}  Using the different tuned values (at least with the latter
having, in general, a different meaning between the two showers), the
predictions on {\it parton level} will differ, though fully simulated, hadronic
events, will yield a comparable description of data. We argue that these
differences should be evaluated carefully, but belong to a future study that
will address the interplay with non-perturbative models in more detail.

\subsection{Simulation Setup}

We consider both parton shower modules available in \hw, the default
angular-ordered shower \cite{Gieseke:2003rz} and the dipole-type shower based
on \cite{Platzer:2009jq,Platzer:2011bc}; in addition to their default
settings, which we have adjusted to make them as similar as possible by
choosing the same $p_\perp$ cutoff and $\alpha_s$ running (the `baseline'
settings for this work), we consider a number of modifications mainly outlined
in Sec.~\ref{sec:scales}, all of which constitute different algorithms in the
sense outlined above. The two showers are very different in their nature: The
angular-ordered, QTilde, shower evolves on the basis of $1\to 2$ splittings
with massive DGLAP functions, using a generalised angular variable and employs
a global recoil scheme once showering has terminated; its available phase
space is intrinsically limited by the angular-ordering criterion, resulting in
a `dead zone', though it is able to generate emissions with transverse momenta
larger than the hard process scale and so typically an additional veto on jet
radiation is imposed (see Sec.~\ref{sec:scales} for more details). The dipole
based shower, Dipoles, uses $2\to 3$ splittings with Catani-Seymour kernels
with an ordering in transverse momentum and so is able to perform recoils on
an emission-by-emission basis; the splitting kernels naturally require the two
possible emitting legs of each dipole to share their phase space and there is
no a priori phase-space limitation, but the available phase space is
controlled by the starting scale of the shower.

\begin{figure}[h]
  \centering
  \includegraphics[scale=0.6]{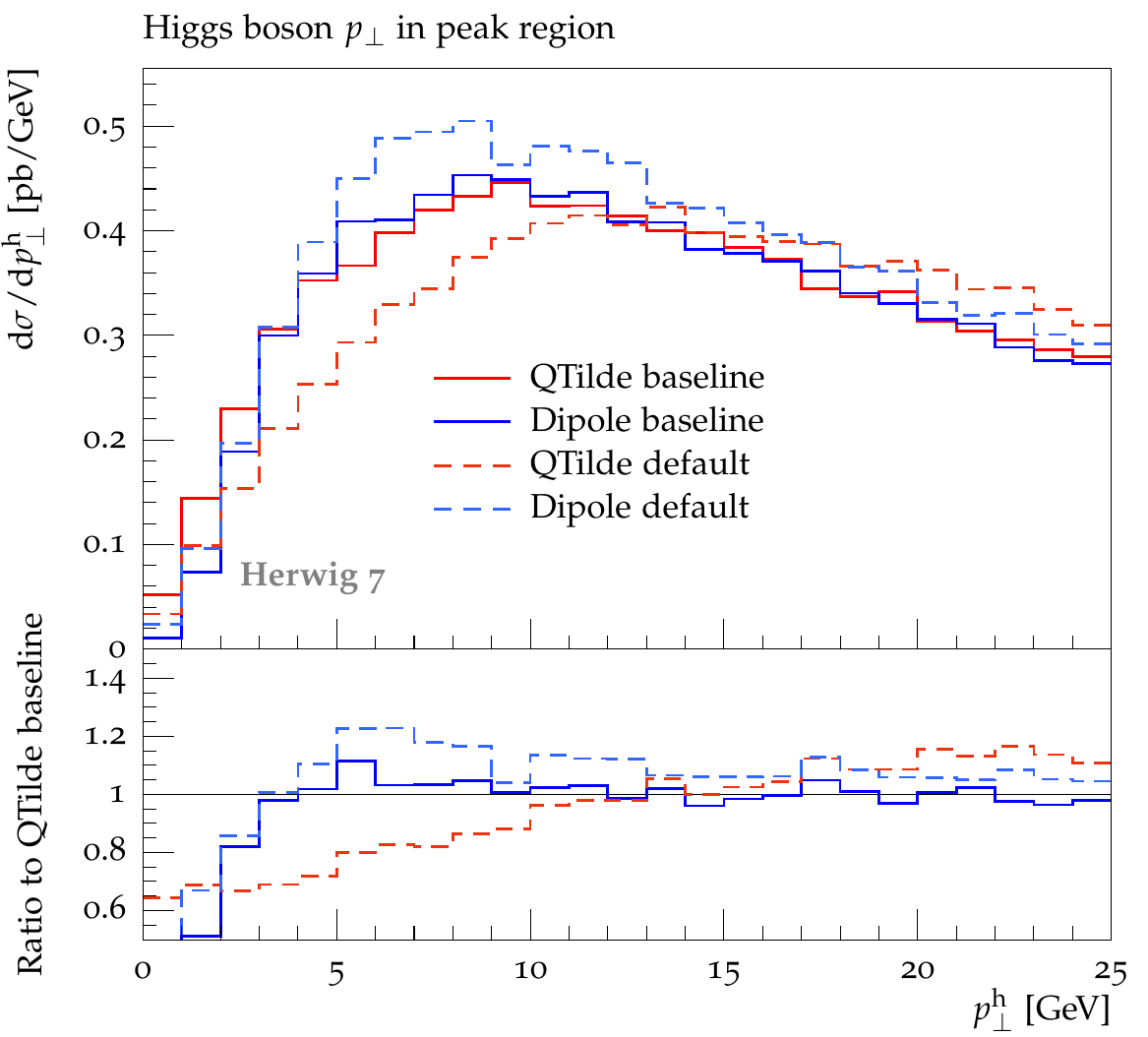}
  \caption{The Higgs boson transverse momentum spectrum comparing the two
    shower algorithms QTilde (red) and Dipole (blue) at their adjusted
    baseline settings used in this comparison.}
  \label{fig:hptbaseline}
\end{figure}

Using the baseline, we find very similar predictions despite the very
different nature of these algorithms.  As an example we show in
Fig.~\ref{fig:hptbaseline} the predictions for the Higgs $p_\perp$ spectrum at
an LHC with $\sqrt{s}=13\ {\rm TeV}$. The only difference between the
algorithms we observe in the very low $p_\perp$ region where the interplay
with the treatment of the remnant and intrinsic transverse momentum smearing
becomes important. For future reference, we have also included results running
the showers at their default settings to highlight what level of interplay
with tuned values and non-perturbative models can be expected.
To be more precise, we use a two-loop running, $\overline{\text{MS}}$,
$\alpha_s$ including CMW correction~\cite{Catani:1990rr} with
$\alpha_s^{CMW}(M_Z)=0.126$ (which corresponds to
$\alpha_s^{\overline{\text{MS}}}(M_Z)=0.118$), and the MMHT2014 NLO PDF set
\cite{Harland-Lang:2014zoa} with five active flavours, interfaced through
\textsf{LHAPDF~6} \cite{Buckley:2014ana}, as far as initial state radiation is
concerned.\footnote{This setup has been chosen such as to later on enable a
  fair comparison to NLO improved simulation that necessitates these orders of
  running.} Hard processes are simulated at leading order (see the previous
discussion), using the \textsf{Matchbox} infrastructure powered by amplitudes
generated by \textsf{MadGraph5\_aMC@NLO} \cite{Alwall:2014hca}. In $e^+e^-$
collissions, we consider di-jet production; at hadron colliders, in addition,
we consider stable $Z$-boson Drell-Yan production, $(e^+e^-j)$ production within the mass window $66~\rm{GeV} < m_{ll} < 116~\rm{GeV}$ around the $Z$ mass,
as well as production of a stable, $125\ {\rm GeV}$, Higgs accompanied by
zero or one jet. In the presence of additional jets in the hard process we use
\textsf{FastJet} \cite{Cacciari:2006sm,Cacciari:2011ma} to perform the
generation cuts; analyses are performed throughout using the \textsf{Rivet}
framework \cite{Buckley:2010ar}, with analysis modules based on existing
experimental and generic Monte Carlo implementations. In $e^+e^-$ collisions,
where we choose a centre of mass energy of $\sqrt{s} = 100$ GeV as baseline,
we reconstruct jets with the Durham algorithm~\cite{Catani:1991hj}, while the
hadron collider setup reconstructs anti-$k_\perp$ jets with a radius of
$R=0.4$ within a rapidity range $|y|<5$ and a transverse momentum threshold of
$p_{\perp} > 20\ {\rm GeV}$.  Parton level without multiple interactions and
hadronisation is employed, and partons up to and including $b$-quarks are
treated as massless objects. Both parton showers mentioned above use a
$p_\perp$ cutoff prescription with a value of $\mu_{\text{IR}}=1\ {\rm
  GeV}$. Electroweak parameters are kept at their default values.

\subsection{Consistency Checks}
\label{subsec:consistency}

The ability to compare different algorithms puts us into the unique position of
performing a number of consistency checks for the uncertainty estimate that we
advocate. In particular, perturbative error bands should cover algorithmic
discrepancies, if these algorithms are expected to deliver the same
accuracy. If that is not the case then the algorithm at hand is
questionable. Furthermore, by construction the shower approximates emissions
in the soft and collinear region. If we force the shower to produce hard
emissions, larger uncertainties are to be expected by a controllable
prescription. Another point is the possibility of double counting hard
emissions. The shower should not cover phase-space regions that are already
covered by the hard process input. This property is typically reflected in
demanding that observables that receive input at fixed order are not significantly
altered by subsequent showering. Clearly, the definition of `region',
which in this case is covered by the veto scale on hard emissions (see
Sec.~\ref{sec:scales} for a more detailed discussion), is again only precise
to the level of accuracy covered by the parton shower and varying this
boundary should serve as a measure of missing logarithmic orders. We emphasise
that a boundary chosen to be far away from the correct ordering behaviour may
introduce severe double counting issues, ultimately impacting on a resummation
of a tower of logarithms which is not typical to the process, {\it i.e.} not
encountered in any higher order corrections to an observable
considered. Furthermore, the perturbative uncertainties for observables in
phase-space regions that do not receive logarithmically enhanced
contributions should be driven by the hard scale alone, while the other scales
have negligible impact. Logarithmically sensitive observables, on the other
hand, should be altered by the parton shower and the uncertainties should be
driven by all possible scale variations together. The setting where this is
least clear is pure jet production, which we will address amongst other
`jetty' processes in Sec.~\ref{sec:jetty}.


\section{Scale Choices, Variations and Profiles}
\label{sec:scales}

\subsection{Phase-Space Restrictions and Profile Choices}

The quantity central to parton showers is the splitting kernel. Its
exponentiation gives rise to the Sudakov form factor, which regulates 
the divergence of the splitting kernel for soft and/or
collinear emissions. On top of this, there are two further crucial
ingredients (besides formally subleading, though not necessarily small
issues like kinematic parametrisations): The evolution variable
chosen, and the phase space accessible at a fixed value of the
evolution variable. Emissions are typically further subject to an
upper bound on their hardness. This cannot be directly deduced from a
priori principles but should be chosen in the order of magnitude of
the typical hardness scale of the process being evolved to avoid the
double counting issues mentioned before.

The central point we are concerned with in this section shall be summarised in
a simplified model of final state radiation. Quite generally, we have to
consider three different scales: a hard scale $K_\perp$ defining the phase
space available to emissions at a fixed transverse momentum; a veto scale
$Q_\perp$ defining the maximum transverse momentum available to emissions; and
the kinematic limit of transverse momentum, $R_\perp$. We consider the
$p_\perp$ spectrum of a single soft emission with splitting kernel (possibly
after an appropriate transformation of the evolution variable into a
transverse momentum)
\begin{multline}
  P_{K_\perp^2}(p^2_\perp,z) = C_i \frac{\alpha_s(p^2_\perp)}{\pi}
  \frac{1}{1-z}\ \times\\
  \theta(z_+(p^2_\perp,K^2_\perp)-z)\theta(z-z_-(p^2_\perp,K^2_\perp))
  \ ,
\end{multline}
where $C_i$ is the colour factor associated with the emitting leg.
The longitudinal momentum fraction, $z$, has limits that read
\begin{equation}
  z_{\pm}(p^2_\perp,K^2_\perp) =
  \frac{1}{2}\left(1\pm\sqrt{1-\frac{p_\perp^2}{K_\perp^2}}\right) \ ,
\end{equation}
in the presence of a hard scale $K_\perp^2$. With emissions weighted
by $\kappa$, an arbitrary function of a veto scale $Q_\perp^2$, we
find a $p_\perp$ spectrum of the form
\begin{multline}
  \frac{{\rm d}{\cal P}}{{\rm d}p_\perp^2 {\rm d}z} = 
  P_{K_\perp^2}(p^2_\perp,z)\frac{\kappa(Q_\perp^2,p_\perp^2)}{p_\perp^2}\ \times\\
  \theta(R_\perp^2-p_\perp^2)\theta(p_\perp^2-\mu_{IR}^2)
  \Delta_{K_\perp^2}(p_\perp^2|Q_\perp^2) \ ,
\end{multline}
with the Sudakov form factor
\begin{multline}
  -\ln \Delta_{K_\perp^2}(p_\perp^2|Q_\perp^2) =\\
  \int_{p_\perp^2}^{R_\perp^2} \frac{{\rm d}q_\perp^2 }{q_\perp^2} 
  \kappa(Q_\perp^2,q_\perp^2)\int {\rm d}z P_{K_\perp^2}(q^2_\perp,z) \ .
\end{multline}
$R_\perp$ denotes the scale that makes all of phase space available to
emissions, while we denote the infrared cutoff by $\mu_{IR}^2$ (we have not shown
the zero $p_\perp$, non-radiating event contribution). Once a hard cutoff
$\kappa(Q_\perp^2,p_\perp^2)=\theta(Q_\perp^2-p_\perp^2)$ is chosen, this
setup is known to reproduce the right anomalous dimensions. It has to be
applied to a full evolution in a hierarchy $Q_\perp^2\to q_\perp^2$ where
$K_\perp^2=Q_\perp^2$ is chosen and the form of the $z$ boundaries being
crucial to produce the correct logarithmic
pattern~\cite{Platzer:2009jq,Hoche:2015sya}. Instead, if one desires to make
all of the phase space available to parton shower emissions,
$K_\perp^2=R_\perp^2$ is chosen and no other than the kinematic constraint
$p_\perp^2<R_\perp^2$ is in place.\footnote{Typically, the splitting kernel
  for exact phase-space factorisation is then accompanied by a damping factor
  $\sim 1 - p_\perp^2/R_\perp^2$ towards the edge of phase space.}

\begin{figure}
  \centering
  \includegraphics[scale=0.6]{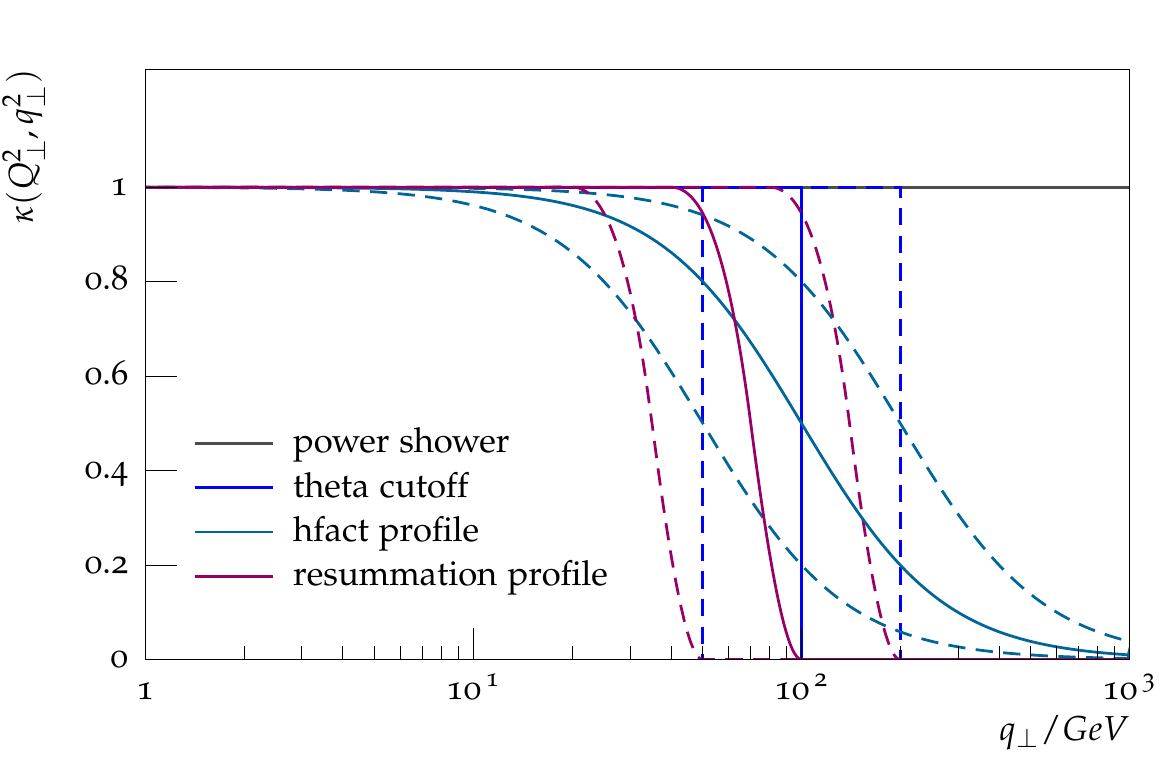}
  \caption{The different profile scale shapes considered in this study at a
    veto scale of $Q_\perp=100\ {\rm GeV}$ (solid) and $Q_\perp=50,200\ {\rm
      GeV}$ (dashed).}
  \label{fig:profilesketch}
\end{figure}

We have here considered the freedom of ensuring suppression of such emissions
by an arbitrary function $\kappa$. We call this weighting function a {\it
  profile scale choice}. One of the subjects of the present study is to
identify sensible profile scale choices; we stress that such a choice is of
{\it algorithmic} nature and not an intrinsic source of uncertainty. We will
consider the following choices, depicted in Fig.~\ref{fig:profilesketch}:
\begin{itemize}
\item \texttt{theta}: $\kappa(Q_\perp^2,q_\perp^2) = \theta(Q_\perp^2
  - q_\perp^2)$, which is expected to reproduce the correct tower of
  logarithms;
\item \texttt{resummation}: $\kappa(Q_\perp^2,q_\perp^2)$ is one below
  $(1-2\rho)\ Q_\perp$, zero above $Q_\perp$, and quadratically interpolating in
  between. This profile is expected to reproduce the correct towers of
  logarithms, and switches off the resummation smoothly towards the hard
  region (currently we use $\rho=0.3$ \footnote{In principle $\rho$ should be
    varied with a reasonable range, though we do not expect a big effect from
    this variation, given the similarities between $\rho=0.3$ and $\rho=0$
    corresponding to the \texttt{theta} profile; see the following
    sections.}):
\begin{multline}
\kappa(Q_\perp^2,q_\perp^2) =\\
\left\{
\begin{array}{ll}
1 & q_\perp/Q_\perp \le 1-2\rho \\
1 - \frac{(1-2\rho-q_\perp/Q_\perp)^2}{2\rho^2} & q_\perp/Q_\perp \in
(1-2\rho,1-\rho] \\
\frac{(1-q_\perp/Q_\perp)^2}{2\rho^2}& q_\perp/Q_\perp \in
(1-\rho,1] \\
0 & q_\perp/Q_\perp > 1
\end{array}
\right. \ ;
\end{multline}
\item \texttt{hfact}: $\kappa(Q_\perp^2,q_\perp^2) =
  \left(1+q_\perp^2/Q_\perp^2\right)^{-1}$, which is also referred to as
  damping factor within the \textsf{POWHEG-BOX}
  implementation\ \cite{Alioli:2010xd}; and
\item \texttt{power} shower: imposing nothing but the phase-space
  restrictions inherent to the shower algorithm considered.
\end{itemize}
Different combinations of $R_\perp^2$ and $K_\perp^2$ can be achieved within
the two showers. In particular, the dipole shower is able to populate the
region up to $K_\perp^2=R_\perp^2$ (`power shower'), while, for $2\to 1$
processes at hadron colliders the angular-ordered phase space, by
construction, imposes $K_\perp^2=Q_\perp^2$ to be the mass of the singlet
which is produced.

The leading logarithmic contribution of the $z$ integration at this
simple qualitative level is given by
\begin{equation}
  \int {\rm d}z P_{K_\perp^2}(q_\perp^2,z) \sim
  \frac{C_i\alpha_s(q_\perp^2)}{\pi}\log\left(\frac{K_\perp^2}{q_\perp^2}\right)
  \ .
\end{equation}
We shall illustrate the impact of the profile scale choice $\kappa$ on the
Sudakov form factor by considering a fixed $\alpha_s$, and evaluate
\begin{eqnarray}
  \label{eqns:logstructure}
  & -&\int_{p_\perp^2}^{R_\perp^2} \frac{{\rm d}q_\perp^2 }{q_\perp^2} 
  \kappa(Q_\perp^2,q_\perp^2)\log\left(\frac{K_\perp^2}{q_\perp^2}\right) =\\\nonumber
&  -&\frac{1}{2}\log^2\left(\frac{K_\perp^2}{p_\perp^2}\right)\kappa(Q_\perp^2,p_\perp^2)\\\nonumber
&  +&\frac{1}{2}\log^2\left(\frac{K_\perp^2}{R_\perp^2}\right)\kappa(Q_\perp^2,R_\perp^2)\\\nonumber
&  +&\frac{1}{2}\int_{p_\perp^2}^{R_\perp^2} {\rm d}q_\perp^2
  \log^2\left(\frac{K_\perp^2}{q_\perp^2}\right)
  \frac{\partial}{\partial q_\perp^2}\kappa(Q_\perp^2,q_\perp^2) \ .
\end{eqnarray}
To obtain the desired resummation properties, namely
\begin{equation}
  \int_{p_\perp^2}^{R_\perp^2} \frac{{\rm d}q_\perp^2 }{q_\perp^2}
  \kappa(Q_\perp^2,q_\perp^2)\log\left(\frac{K_\perp^2}{q_\perp^2}\right)
  \sim \frac{1}{2}\log^2\left(\frac{Q_\perp^2}{p_\perp^2}\right) \ ,
\end{equation}
a number of limitations on $\kappa$ and the other scale choices need
to be imposed. Clearly, the limiting cases for small and large
transverse momenta need to be reproduced;
\begin{eqnarray}
  \kappa(Q_\perp^2,p_\perp^2) \to 1 &\quad& p_\perp^2 \ll Q_\perp^2\ ,\\\nonumber
  \kappa(Q_\perp^2,p_\perp^2) \to 0 &\quad& q_\perp^2\sim R_\perp^2 \gg Q_\perp^2 \ .
\end{eqnarray}
While this is the case for all of the profiles we considered in this
study, it is not sufficient to produce the desired tower of logarithms.
Imposing the former restriction we still require that:
\begin{itemize}
\item $K_\perp^2\sim Q_\perp^2$ is imposed by the $z$ boundaries; {\it
  and}
\item $\kappa(Q_\perp^2,q_\perp^2) \sim \text{const}$ whenever
  $q_\perp^2$ is not of the order of $Q_\perp^2$ for the term
  involving the derivative of $\kappa$ to become subleading.
\end{itemize}
Specifically the first restriction is only guaranteed by either the angular-ordered 
phase space which naturally imposes this restriction, or the
restricted phase space chosen for the dipole shower. The second restriction
also excludes choices of $\kappa$ providing a ratio of logarithms to
effectively replace $K_\perp^2$ by $Q_\perp^2$ in the first term in
Eq.~\ref{eqns:logstructure}. To this extent, we conclude that only those
profiles that are {\it narrow} smeared versions (in the sense of varying only
in a region where $Q_\perp^2/q_\perp^2$ is of order one) of a
\texttt{theta}-type cutoff will provide the proper tower of
logarithms. Choices such as the \texttt{resummation} profile are desirable to
avoid discontinuities introduced by the \texttt{theta}-type cutoff which are
beyond the accuracy considered, while keeping the resummation
properties of the parton shower; the profile we consider here is only one such kind,
and there is no restriction on the exact form considered. The name
`profile' is chosen since the treatment of the hard scale we consider here
closely resembles prescriptions on scale variations within the analytic
resummation context \cite{Tackmann:Profiles}.

\subsection{Identifying a `Resummation Scale'}

The hard veto scale $Q_\perp^2$ is the scale that, when considering
transverse momentum spectra as outlined in the previous section, is closest in
role to a resummation scale in analytical resummation. However, it is not
typically the same as a shower starting scale. Though in our case this
statement is true for the dipole shower, no such notion exists for the angular-ordered
shower where typically the masses of the emitting dipoles set the
shower starting scale owing to the angular-ordered phase space. In the latter
case, emissions exceeding the transverse momenta of jets present in the hard
process are possible and an additional veto on transverse momenta generated by
the shower is applied; the value of this veto is, in this case, the analogue
of the starting scale of the dipole shower. The resummation scale of the
typical $q_\perp$-resummation can thus not directly be related to an analogue
hard scale present in different shower algorithms, especially when they evolve
in a variable different from the transverse momentum and hence built up the
full spectrum from multiple, differently ordered emissions.

For both the showers considered here, transverse momenta of parton shower
emissions are expected to be limited or suppressed by the scale $Q_\perp^2$,
which on very general grounds should thus be of the order of a typical scale
of the hard process, {\it i.e.} the factorisation scale. As with fixed-order
calculations the residual dependence on $Q_\perp^2$ is expected to become
smaller as more and more logarithmic orders are incorporated. This implies a
pattern of scale compensation through successive logarithmic orders, which a
parton shower can typically only guarantee at the level of at {\it most}
next-to-leading logarithms (NLL).

\subsection{Scale Variations}
\label{subsec:scales}
Having chosen a reasonable profile scale and value of $K_\perp^2\sim
Q_\perp^2$, the leading behaviour of the Sudakov exponent takes the well-known
form
\begin{multline}
-\ln \Delta(p_\perp^2|Q_\perp^2) =\\
\int_{p_\perp^2}^{Q_\perp^2} \frac{{\rm
    d}q^2}{q^2}\left( A(\alpha_s(q^2))\ln\left(\frac{Q_\perp^2}{q^2}\right) +
B(\alpha_s(q^2)) \right) \ ,
\end{multline}
where the highest level of accuracy one can hope for with coherent evolution
is NLL accuracy, neglecting subleading colour correlations, at least for
some observables and typically limited phase-space regions
\cite{Catani:1990rr},
\begin{equation}
A(\alpha_s) =
\frac{C_i}{2}\frac{\alpha_s}{\pi}\left(1+\frac{K_g}{2}\frac{\alpha_s}{\pi}
\right)\qquad B(\alpha_s) = \frac{\alpha_s}{\pi} \frac{\gamma_i}{2}
\end{equation}
along with a two-loop running of $\alpha_s$. 

In addition to variations of the scales in the hard process, $\mu'_{R/F}=\xi_H
\mu_{R/F}$, we vary both the hard veto scale, $\mu_Q=\xi_Q Q_{\perp}$, and the
arguments of $\alpha_s$ and the PDFs in the parton shower splitting kernels,
$\mu_S=\xi_S q_\perp$. We constrain the number of possible variations to be
$\xi \in [1/2,1,2]$. This spans a cube of, $\xi_H \otimes \xi_S \otimes
\xi_Q$, 27 combinations. All these choices are connected to logarithmic scale
choices. There is therefore no a priori way of reducing their number. We
emphasise that in principle only the full 27-point envelope constitutes a
comprehensive uncertainty measure. We therefore always produce the full
envelope along with envelopes for each of the individual variations. Using
this it is possible to observe which scale drives the overall uncertainty in a
particular region of phase space. While one clearly expects the variation of
$Q_\perp^2$ to cancel out to the level of NLL accuracy~\cite{Hamilton:Scales}
(if this is indeed resembled by the parton shower), the situation is less
clear for the other variation and different proposals have been made as to
what extent the contribution at the level of NLL contributions should be
canceled (see {\it e.g.}  \cite{Dasgupta:2001eq} for a discussion) or
otherwise considered as a probe of where precisely higher accuracy of the
shower is missing. We do not consider introducing any terms that cancel these
variations to the NLL order, and postpone a detailed analysis of this issue to
future work. We do, however, analyse these variations as we are convinced that
they are another clean handle on {\it controlling} where we expect,
specifically, soft emissions and contributions by the hadronisation model to
dominate. A recent Les Houches study \cite{LesHouches:2015} has also shown
that, when not taking into account the full variations of this kind,
discrepancies between different shower algorithms, which are expected to be
similar, are not covered within these variations.

\subsection{Real-life Constraints}

Besides the unclear definition of a resummation scale in the context of
different shower algorithms, another word of caution needs to be raised when
considering the hard veto scales: There are cases in which there is no
meaningful variation as the hard scale is a fixed quantity such as the mass
of an independently evolving final state emitting system, {\it e.g.} showers
in $e^+e^-\to$ hadrons. It is not clear how one would quantify the respective
shower uncertainty in this case, besides looking at shape differences
encountered at different centre-of-mass energies of the $e^+e^-$ collider to
quantify the scaling of the predictions with respect to ratios of the hard
scale to the infrared sensitive quantity considered. Already this observation
clearly marks the fact that no claim of a full and well-understood uncertainty
recipe can be made at this point, but only are we able to perform initial
steps in this direction. Similarly for the \texttt{power} shower there is no
meaningful variation of $\mu_Q$. It can also happen, as is the case for the
angular-ordered shower, that the algorithm chosen naturally imposes an upper
bound on the hardness of the emission. In the case of Drell-Yan type
processes, the angular-ordered shower, for example, will only allow for a
down-variation of $Q_\perp^2$ and is thus questionable as to whether this variation
in these cases is the right measure.

As with the small scales probed in the evolution of the parton shower,
$\mu_{R,F}$ variations in the shower may actually encounter regions where typically
some cutoff or freezing-like behaviour is imposed to both, the running of
$\alpha_s$ and the parton distributions functions, which may result in
interesting dynamics when variation of such small scales is used to infer
uncertainties -- a variation of the freezing prescription may thus be
desirable, as well.

\section{Clean Benchmarks}
\label{sec:clean}
To begin exploring the uncertainties that arise from the considerations of
Sec.~\ref{sec:scales} we start by studying `clean benchmarks', {\it i.e.}
hard processes with the least number of legs: $e^+e^-$
annihilation, and Drell-Yan-type $2\to 1$ processes producing either a $Z$ or
Higgs boson. For the case of $e^+e^-$ collisions, the notion of a hard veto
scale does not directly exist owing to the fact that the phase-space boundary
and relevant hard scale coincide. However, we can compare variations of the
collision energy and quantify this impact at the level of normalised
distributions to acquire a handle on variations of the logarithmic structure
similar to hadron-hadron collisions\footnote{We do not consider deep
  inelastic scattering, which is interesting in its own respect. Similarly, a
  (hypothetical) $e^+e^-\to gg$ collider setting should be explored to
  complement our studies of $Z$ versus $H$ production in $pp$ collisions; we
  postpone these discussions to later work elaborating on the interplay with
  hadronizsation models, where these differences are expected to be more
  relevant; the reader is also referred to the Les Houches study
  \cite{LesHouches:2015} in this context.}.  On top of the three scales
$\mu_{H,S,Q}$ described above, we vary the infrared cutoff of the shower by a
factor of $1/2$ and $2$ for the $e^+e^-$ setting, in order to obtain a first
indication of how much dynamics of the shower is expected to be absorbed into
hadronisation effects; notice that varying the argument of $\alpha_s$ may
serve a similar purpose.

\subsection{Final State Showers}
$e^+ e^- \to qq$ provides the clean environment to study final state
radiation. Note that in this case the \texttt{power} and
\texttt{theta} profile coincide, which is also our choice in the
following.
\begin{figure}[t]
    \centering
    \includegraphics[width=0.4\textwidth]{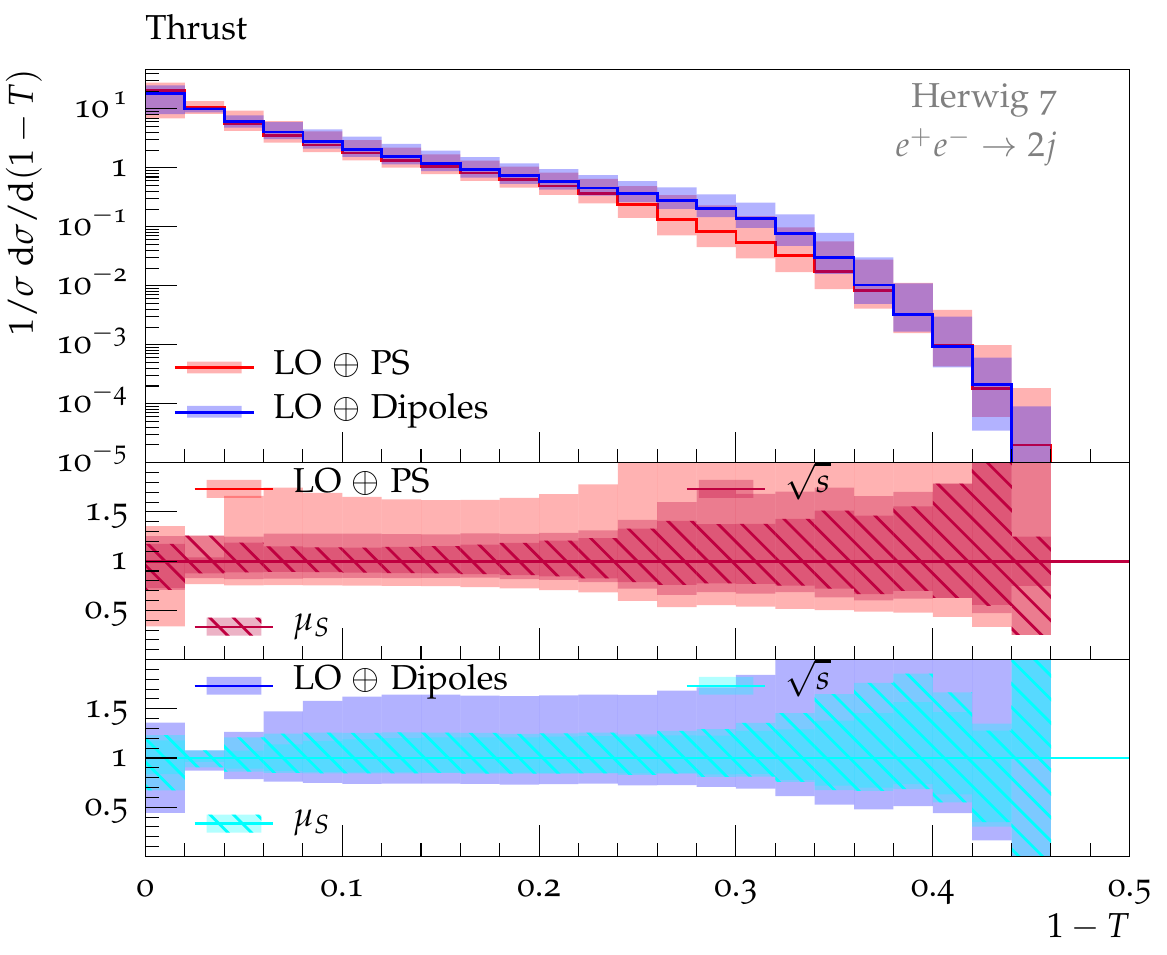}
    \caption{Thrust distribution for the QTilde shower (red) and Dipole 
      shower (blue). The main plot envelopes consist of the variations
      $\sqrt{s}\otimes\mu_S$. The ratio plot for each individual shower contains
      the contributions from the individual variation of each scale which are shown
      relative to the full envelope (ratio, top-left).}
    \label{fig:ee:thrust}
\end{figure}

The Thrust distribution, Fig.~\ref{fig:ee:thrust}, shows
good agreement between showers; this is true both for the central
prediction and its variations, and shows that they possess the same
resummation accuracy. Differences that do emerge between the showers
are related to cutoff effects and non-radiating events in the region
towards $T=1$; these offer no insight into the resummation
properties. A further difference emerges from the dead-zone of the
QTilde shower, however this is a region that can be supplemented by
using matching or ME corrections. For this observable we note that the
$\sqrt{s}$ and $\mu_S$ variations are similar in magnitude.
\begin{figure}[!h]
    \centering
    \includegraphics[width=0.4\textwidth]{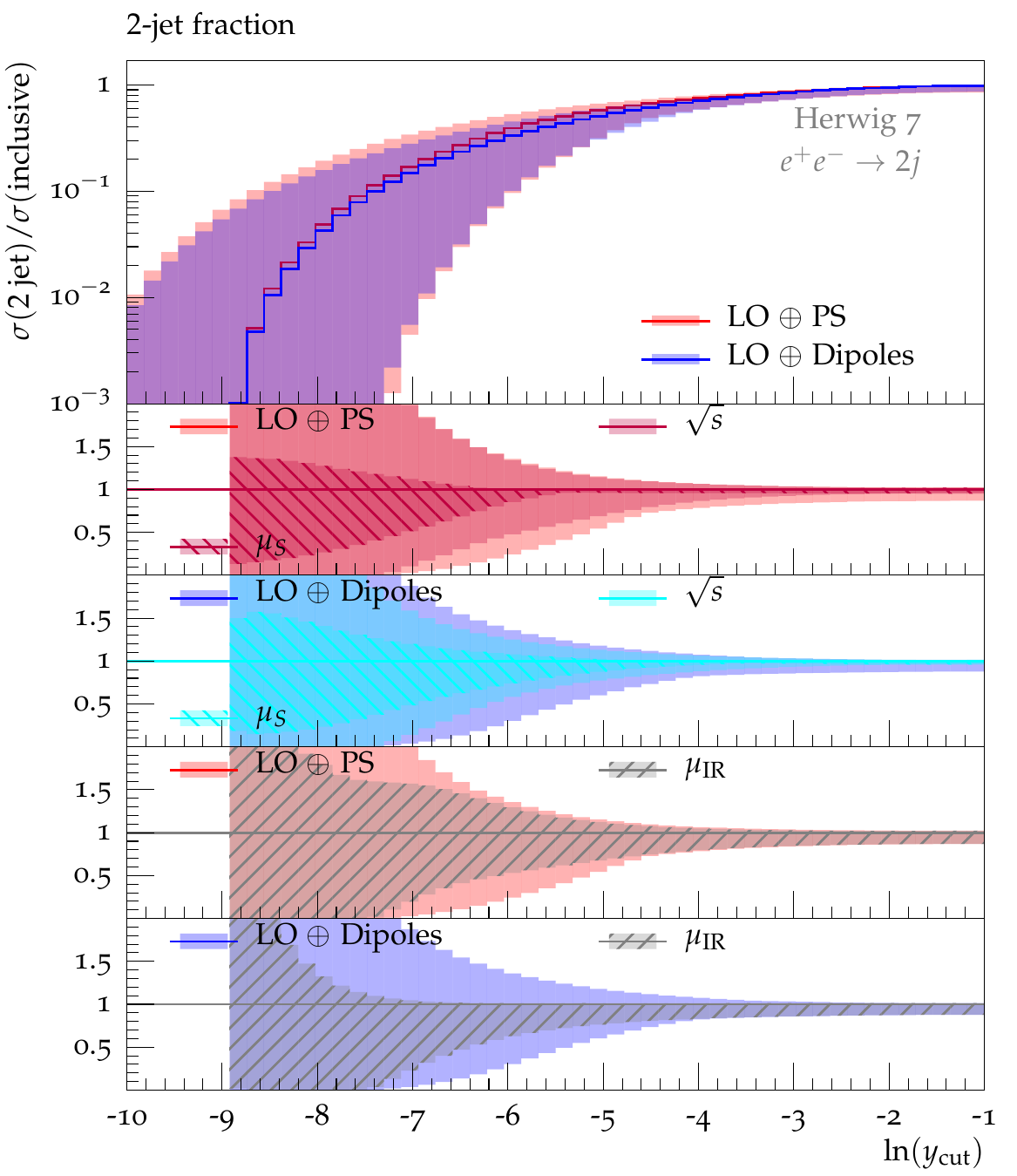}
    \caption{Distribution of the 2-jet fraction against $y_\mathrm{cut}$. The
      main plot contains the envelope of the scale variations $\sqrt{s}\otimes
      \mu_S$ for the QTilde shower (red) and Dipole shower (blue). The first
      two ratio plots contain the contributions from the individual variation of
      each scale relative to the full envelope (ratio, top-left). The last two
      ratio plots show the $\mu_\mathrm{IR}$ variations (ratio, top-right), and,
      for size comparison, we also plot the envelopes of the main plot (ratio,
      top-left).}
    \label{fig:ee:r2}
\end{figure}

In Fig.~\ref{fig:ee:r2} we show results for the integrated two-jet rate; the
uncertainties are dominated by $\sqrt{s}$ as well as cutoff variations at
small $y_\mathrm{cut}$. Again, the overall uncertainties are comparable
between the showers; as expected, we obtain large uncertainties in the small $y_{\rm cut}$ region,
which is dominated by hadronisation effects.
\begin{figure}[t!]
    \centering
    \includegraphics[width=0.4\textwidth]{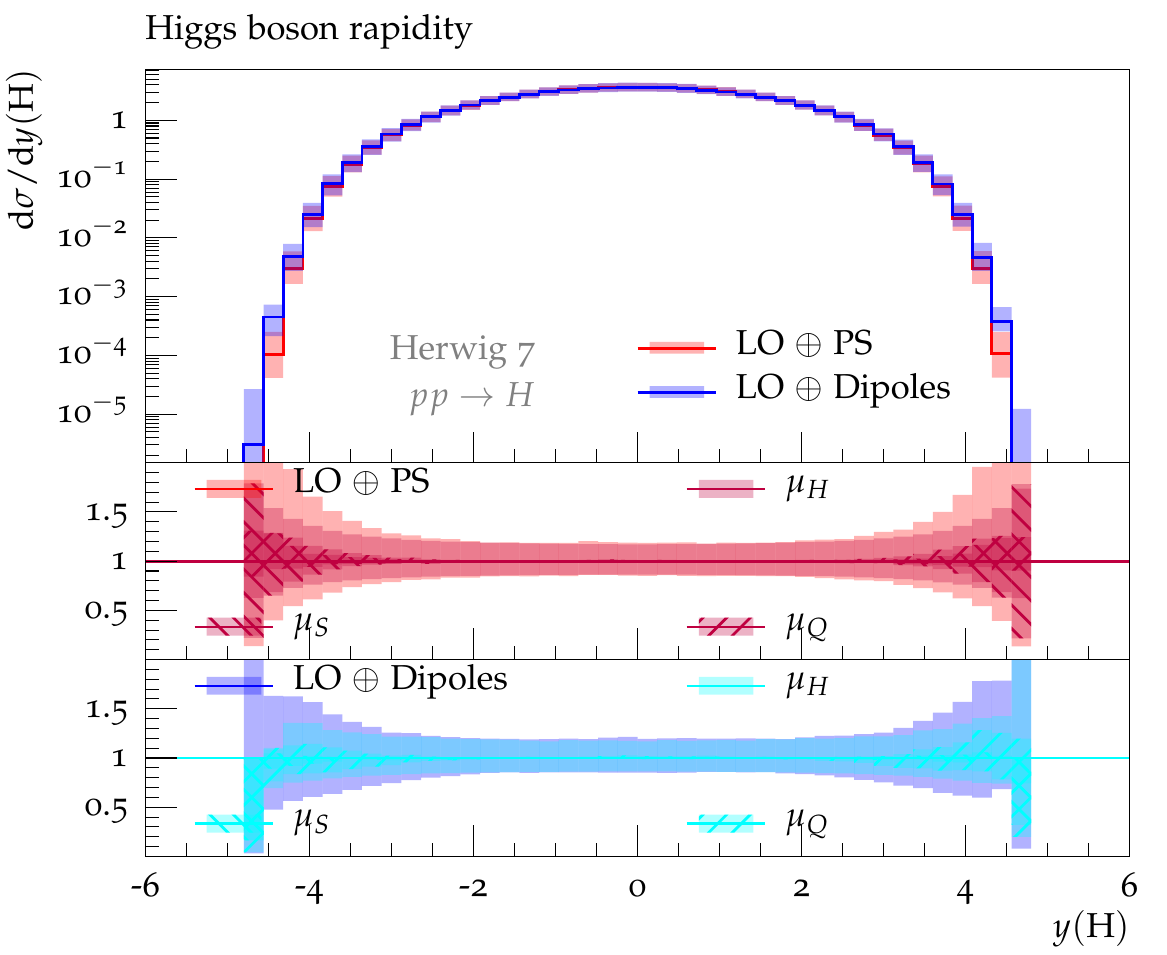}
    \caption{Rapidity of the Higgs boson for both the QTilde shower (red) and the 
      Dipole shower (blue). The main plot envelope contains the full scale variations
      for each shower, and the ratio plots contain their breakdown in terms of the individual 
      scales relative to the full envelope (ratio, top-left).
      The lines shown are for the \texttt{resummation} profile.}
    \label{fig:pp:mu:H_y}
\end{figure}
\begin{figure}[t!]
    \centering
    \includegraphics[width=0.4\textwidth]{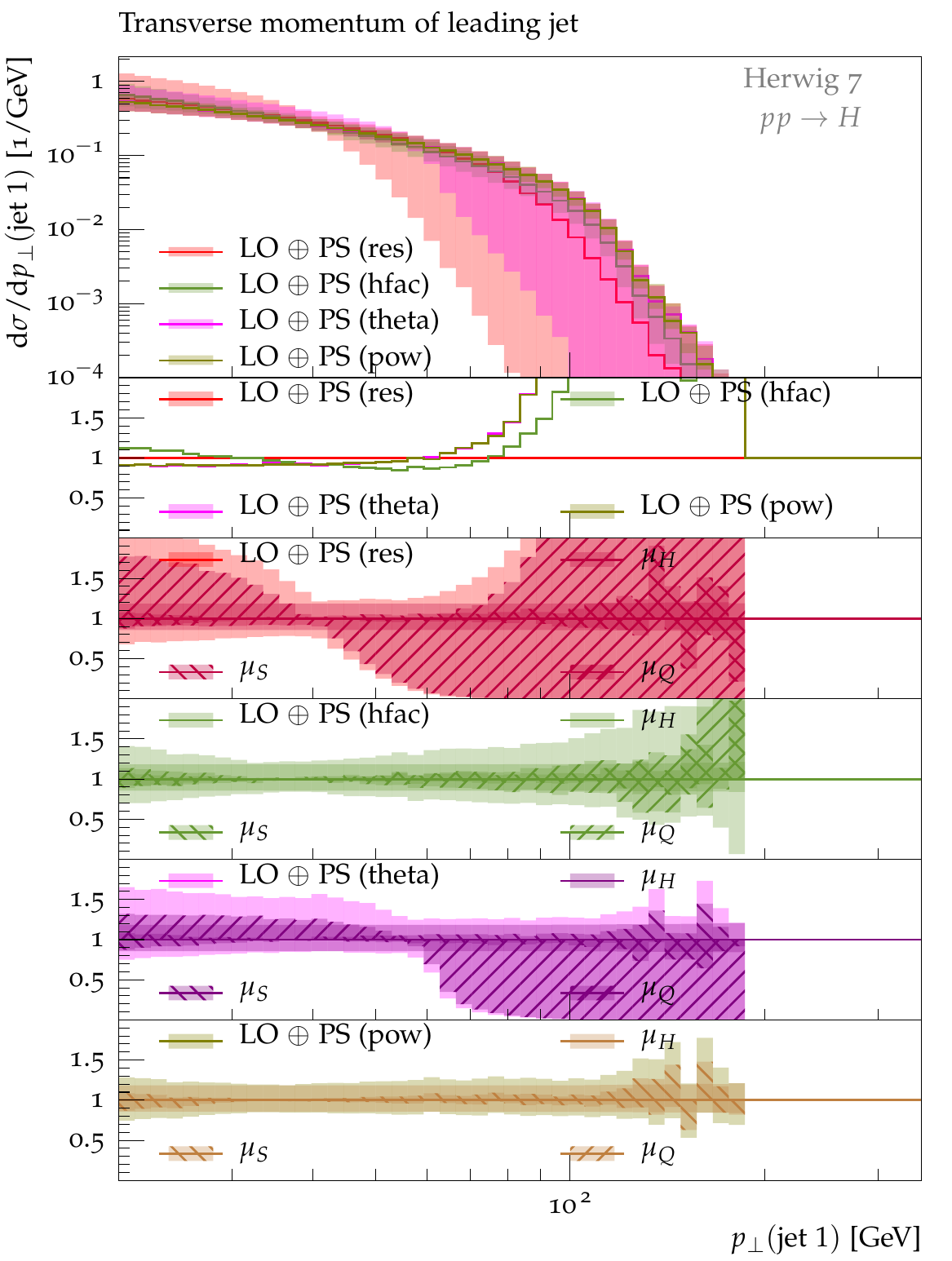}
    \caption{Transverse momentum of the leading jet in Higgs production with the
      QTilde shower. The main plot envelopes consist of the full set of 
      $\mu_H,\mu_S,\mu_Q$ variations for the \texttt{resummation} (red),
      \texttt{hfact} (green), \texttt{theta} (pink), \texttt{pow} (brown).
      The first ratio plot shows the central predictions for each profile relative to the \texttt{resummation} profile. The subsequent ratio plots show the variations of individual scales relative to the full envelope for each profile (ratio, top-left).}
    \label{fig:pp:mu:H:j1_pt_qt}
\end{figure}
\begin{figure}[t!]
    \centering
    \includegraphics[width=0.4\textwidth]{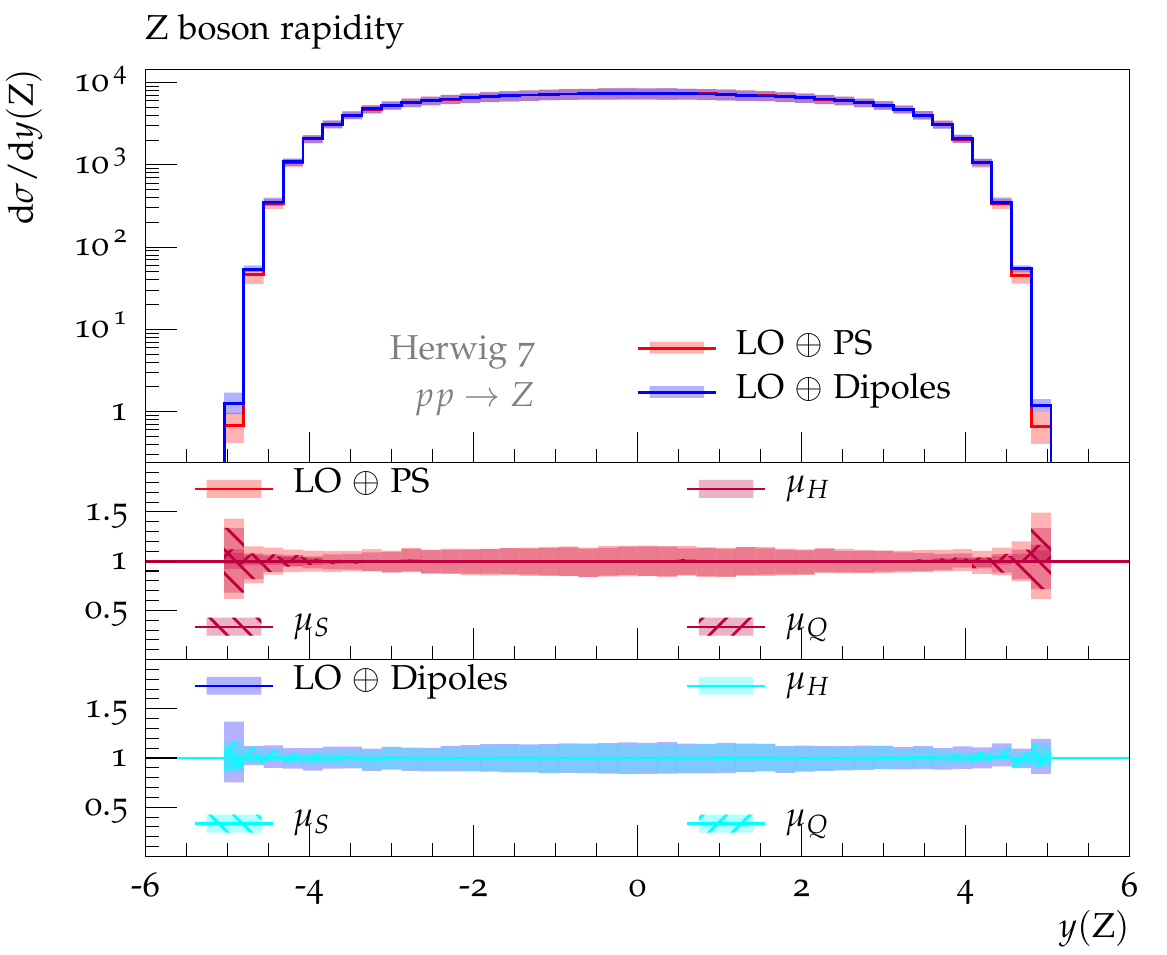}
    \caption{Rapidity of the Z boson for both the QTilde shower (red) and the 
      Dipole shower (blue). The main plot envelope contains the full scale variations
      for each shower, and the ratio plots contain their breakdown in terms of the individual 
      scales relative to the full envelope (ratio, top-left).
      The lines shown are for the \texttt{resummation} profile.}
    \label{fig:pp:mu:Z_y}
\end{figure}
\begin{figure}[t!]
  \centering
  \includegraphics[width=0.4\textwidth]{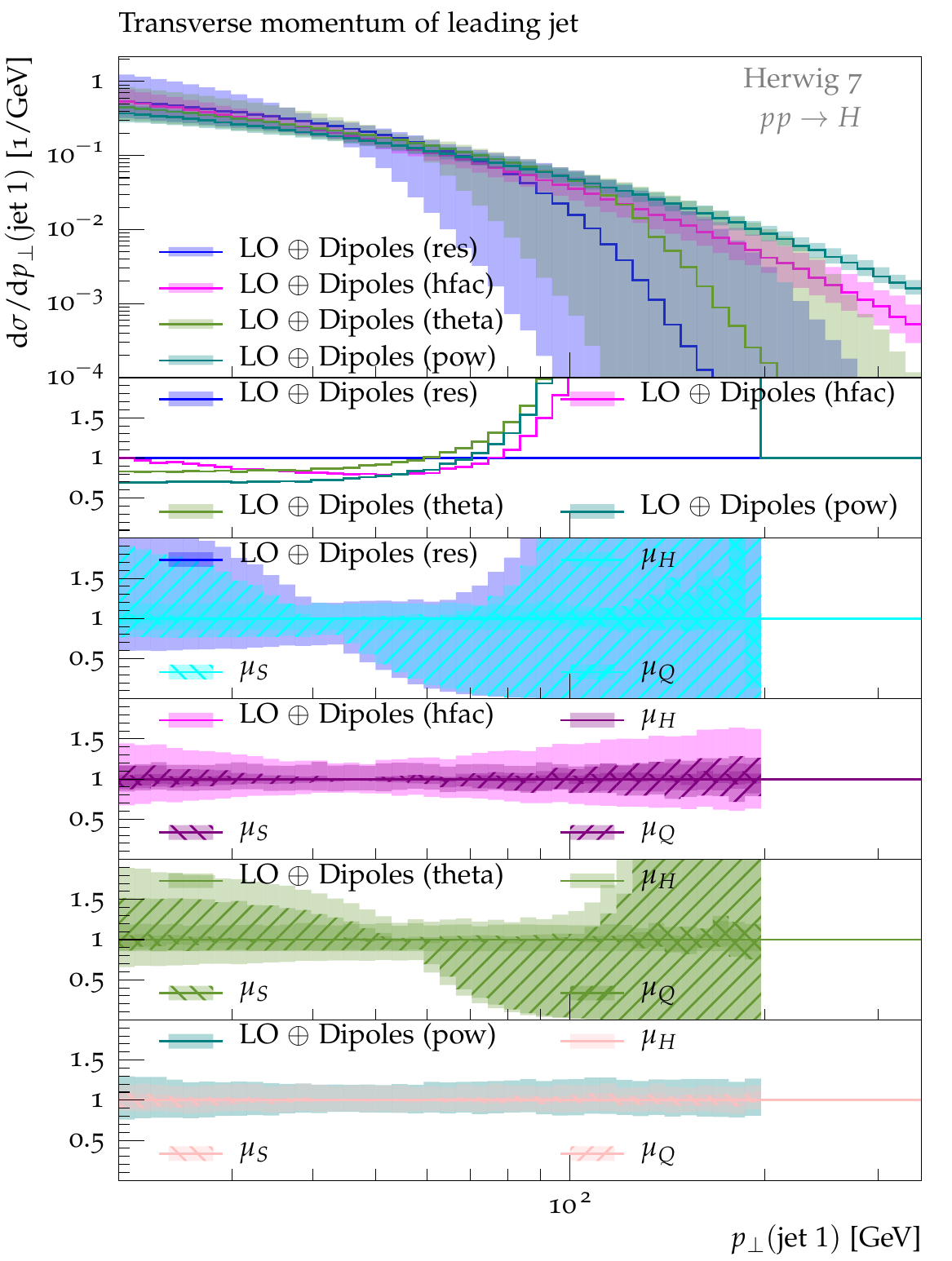}
  \caption{
      Transverse momentum of the leading jet in Higgs production with the
      Dipole shower. The main plot envelopes consist of the full set of 
      $\mu_H,\mu_S,\mu_Q$ variations for the  \texttt{resummation} (blue),
      \texttt{hfact} (pink), \texttt{theta} (green), \texttt{pow} (teal).
      The first ratio plot shows the central predictions for each profile
      relative to the \texttt{resummation} profile. The subsequent ratio 
      plots show the variations of individual scales relative to the full
      envelope for each profile (ratio, top-left).}
    \label{fig:pp:mu:H:j1_pt_dip}
\end{figure}

\subsection{Initial State Showers}

As far as initial-state showering is concerned, we investigate a gluon-initiated
process $pp \to H$ (in the large-$m_t$ effective theory), and a
quark-initiated process $pp \to Z$; these particles are set stable for
simplicity.
Inclusive observables, such as the rapidity of the resonance in this case, are
quantities expected to be well described by the matrix element, and thus
should be unmodified by the parton shower; this is reflected in
Figs.~\ref{fig:pp:mu:H_y} and \ref{fig:pp:mu:Z_y} where both showers display
good agreement, with uncertainties mainly driven by the hard process
variation. The differences in magnitude should be attributed to different
couplings for each process, with envelope shape differences attributed to the
PDFs.

The jets in these samples are generated solely from the parton shower;
therefore the $p_\perp$ of the leading (hardest) jet directly probes
the impact of the profile scales.
\begin{figure}[t]
    \centering
    \includegraphics[width=0.4\textwidth]{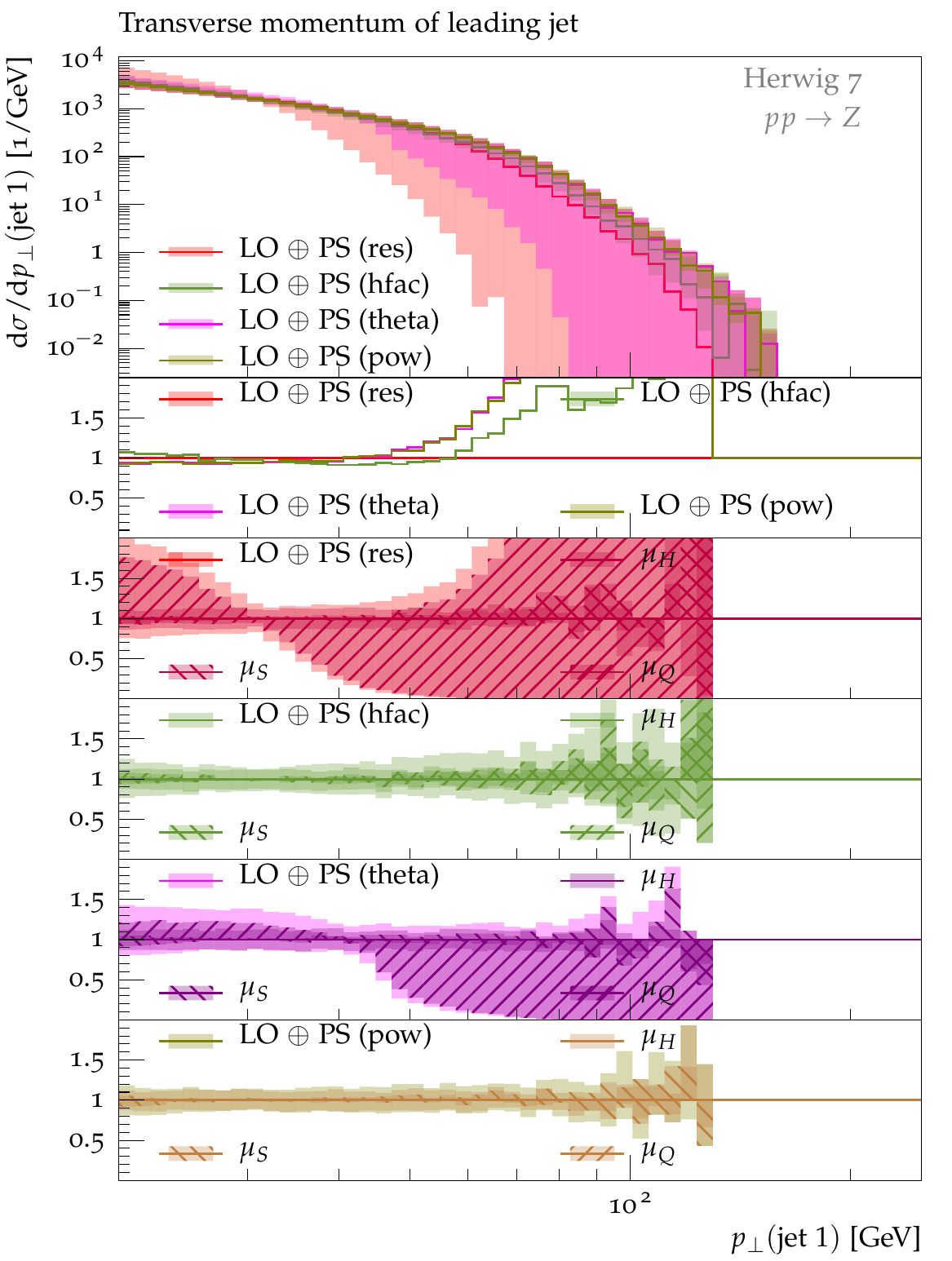}
    \caption{Transverse momentum of the leading jet in Z production with the
      QTilde shower. The main plot envelopes consist of the full set of 
      $\mu_H,\mu_S,\mu_Q$ variations for the \texttt{resummation} (red),
      \texttt{hfact} (green), \texttt{theta} (pink), \texttt{pow} (brown).
      The first ratio plot shows the central predictions for each profile
      relative to the \texttt{resummation} profile. The subsequent ratio 
      plots show the variations of individual scales relative to the full
      envelope for each profile (ratio, top-left).}
    \label{fig:pp:mu:Z:j1_pt_qt}
\end{figure}
\begin{figure}[t]
  \centering
  \includegraphics[width=0.4\textwidth]{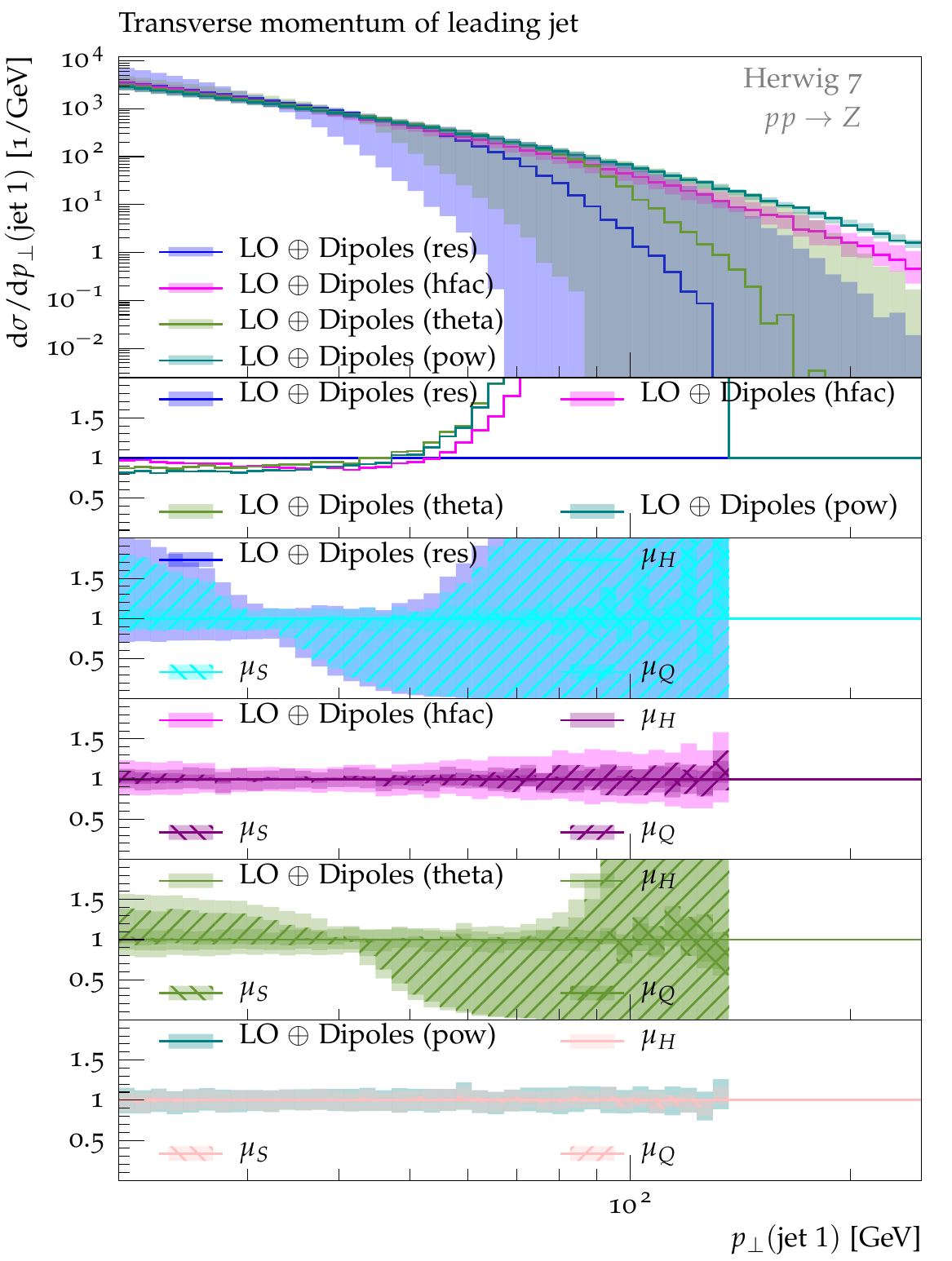}
  \caption{
  Transverse momentum of the leading jet in Z production with the
      Dipole shower. The main plot envelopes consist of the full set of 
      $\mu_H,\mu_S,\mu_Q$ variations for the \texttt{resummation} (blue),
      \texttt{hfact} (pink), \texttt{theta} (green), \texttt{pow} (teal).
      The first ratio plot shows the central predictions for each profile
      relative to the \texttt{resummation} profile. The subsequent ratio 
      plots show the variations of individual scales relative to the full
      envelope for each profile (ratio, top-left).}
    \label{fig:pp:mu:Z:j1_pt_dip}
\end{figure}

Comparing Figs.~\ref{fig:pp:mu:H:j1_pt_qt}-\ref{fig:pp:mu:Z:j1_pt_dip}, we
find that the different profile choices exhibit significantly different
behaviours, both amongst themselves as well as between different showers. The
 \texttt{resummation}~and \texttt{theta} profiles, as intended, yield
comparable results in terms of central predictions and uncertainties and
across the different shower algorithms. This clearly shows that we can indeed
expect the same resummation accuracy using these profiles. The variations
towards high $p_\perp$ for the \texttt{theta} profile expose the effect of the
different phase-space limitations. In the QTilde shower the upward variation
of the scales ($\mu_Q$) is ultimately irrelevant, as there are no possible
emissions at this scale; looking at the dipole shower one sees the effect of
such variations.  However, this is not the case for the \texttt{resummation}
profile whose interpolating region is sensitive to such variations, and
displays similar variations between showers.

For large transverse momenta, the uncertainties should reflect the
case that parton-shower emissions in these regions are unreliable. We
observe this for both the \texttt{theta} and \texttt{resummation}
profiles and to some extent for the \texttt{hfact} choice, though the
variation is considerably smaller than indicated by the
\texttt{theta}-type choices. The \texttt{power} shower, however, shows
no increased uncertainty and in fact is dominated by variations of
$\mu_H$. Given the marked differences in the hardness of jets between
the two showers, the \texttt{power} shower seems to offer no handle
towards the assessment of shower uncertainties. We can also clearly
observe the intrinsic limitation of the QTilde shower phase space,
that in this case is not able to populate high-$p_\perp$ emissions
which ultimately needs to be supplied by matching and/or matrix
element corrections similarly to the `dead zone' effect in $e^+e^-$
collisions.

We therefore conclude that within this basic setting the showers and profile
scale choices do admit the expected behaviour, and the two showers using
\texttt{theta}-type profiles exhibit similar central predictions and
uncertainties.

\section{Jetty Processes}
\label{sec:jetty}

Having established shower uncertainties using simple benchmark processes, the
next simplest examples are the processes studied in Sec.~\ref{sec:clean} with
an additional hard emission off the hard process, e.g. $H/Z$ plus one
(inclusive) jet. In addition, pure di-jet production is investigated because
of the absence of a colour singlet setting a hard scale and the related
ambiguities in possible hard scale choices. We do not investigate the shower
cutoff as we shall now focus on properties which are not expected to be
significantly altered by hadronisation effects.
\begin{figure}[t]
\centering
\includegraphics[width=0.4\textwidth]{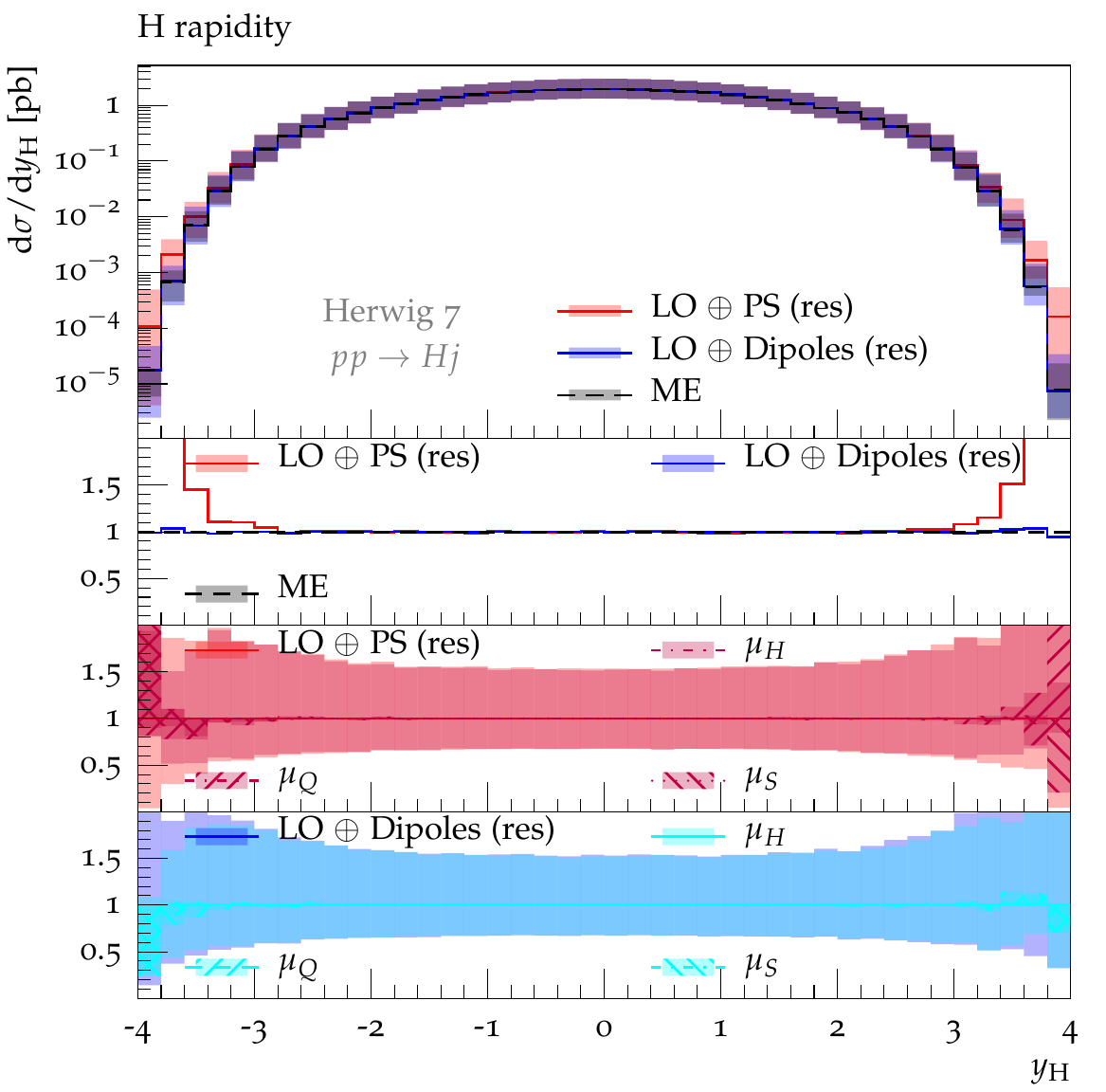}
\caption{Rapidity of the Higgs in Higgs plus one jet events for the QTilde
  (red) and Dipole (blue) shower compared to the matrix element prediction
  (black). We show results obtained with the resummation profile. The error
  bands are computed from all allowed scale choices (see text). Top ratio
  plot: QTilde vs. Dipole and ME. Second ratio plot: QTilde with full error
  band vs. variation of only either $\mu_H$, $\mu_Q$ or $\mu_S$. Third ratio
  plot: Dipole with full error band vs. variation of only either $\mu_H$,
  $\mu_Q$ or $\mu_S$. }
\label{fig:H_rap}
\end{figure}
\begin{figure}[t]
\centering
\includegraphics[width=0.4\textwidth]{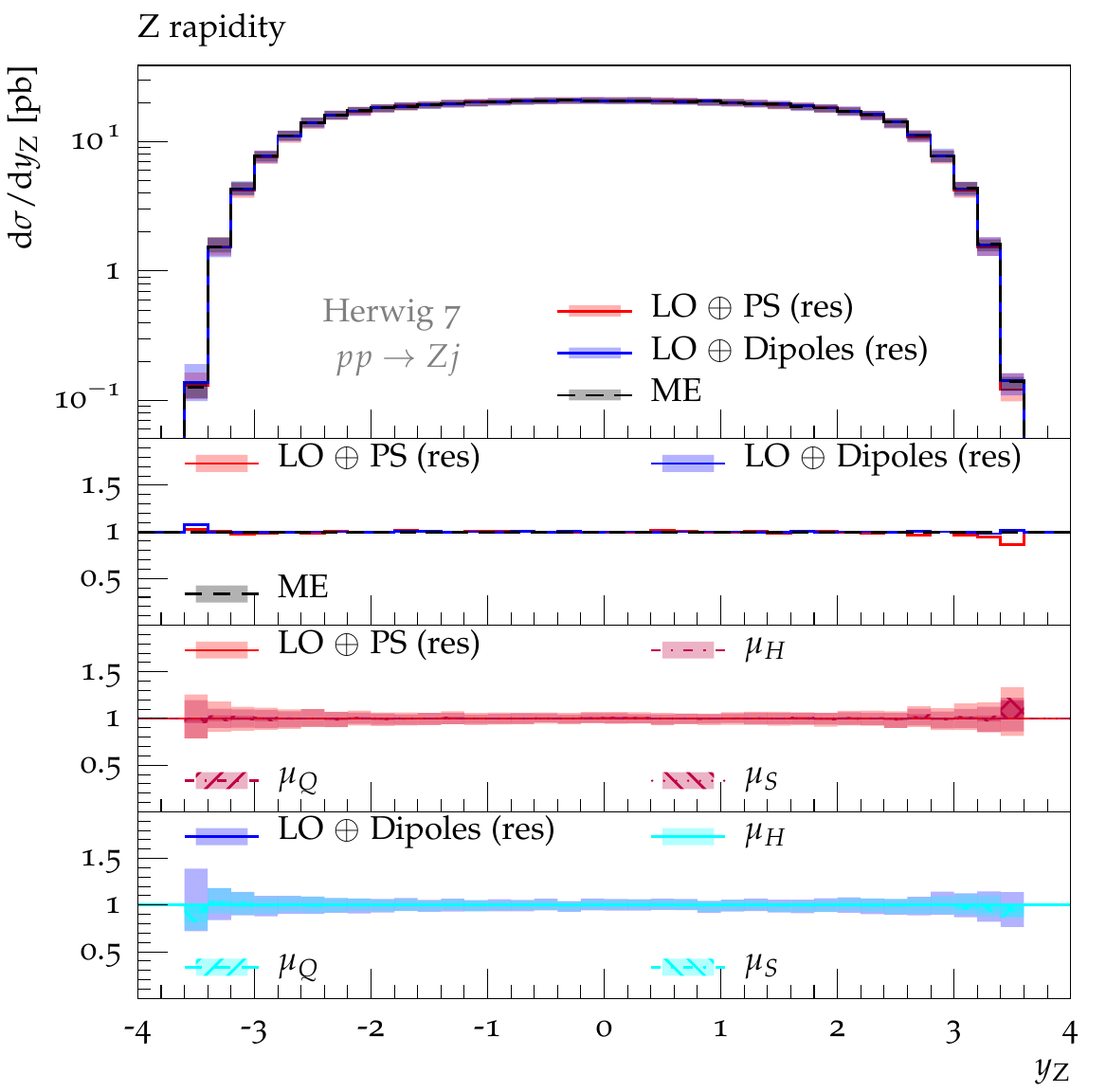}
\caption{Rapidity of the Z in Z plus one jet events for the QTilde (red) and
  Dipole (blue) shower compared to the matrix element prediction (black). We
  show results obtained with the resummation profile. The error bands are
  computed from all allowed scale choices (see text). Top ratio plot: QTilde
  vs. Dipole and ME. Second ratio plot: QTilde with full error band
  vs. variation of only either $\mu_H$, $\mu_Q$ or $\mu_S$. Third ratio plot:
  Dipole with full error band vs. variation of only either $\mu_H$, $\mu_Q$ or
  $\mu_S$.}
\label{fig:Z_rap}
\end{figure}

As with the clean benchmarks presented in Sec.~\ref{sec:clean}, we
consider variations of the three relevant scales discussed in
Sec.~\ref{sec:scales}, changing them by factors $1/2$ and $2$,
respectively, to span a cube of a total of $27$ variations; we will
also perform cross-validations between both available showers. From
arguments given in Sec.~\ref{sec:scales} we expect observables and/or
regions in phase space where the uncertainty is mainly driven by
$\xi_\text{H}$, {\it i.e.} in the case of inclusive observables. As
all uncertainties connected with scale choices stem from logarithmic
arguments there is no a priori way to exclude any of the possible
variations when determining shower uncertainties, unless one is able
to identify scale compensation patterns between the different scales
for which we see no evidence in the setting considered in this study.

For the rapidity distributions of the Higgs and $Z$ boson, shown in
Fig.~\ref{fig:H_rap} and~\ref{fig:Z_rap}, respectively, we find that the
distributions are consistent with the prediction of the hard matrix element, as
is expected from such inclusive quantities; this applies to all of the profile
scales considered, with the \texttt{power} shower showing larger deviations in
the forward region. Scale variations affect these observable mainly through
variations present in the hard process.

\begin{figure}[t!]
\centering
\includegraphics[width=0.4\textwidth]{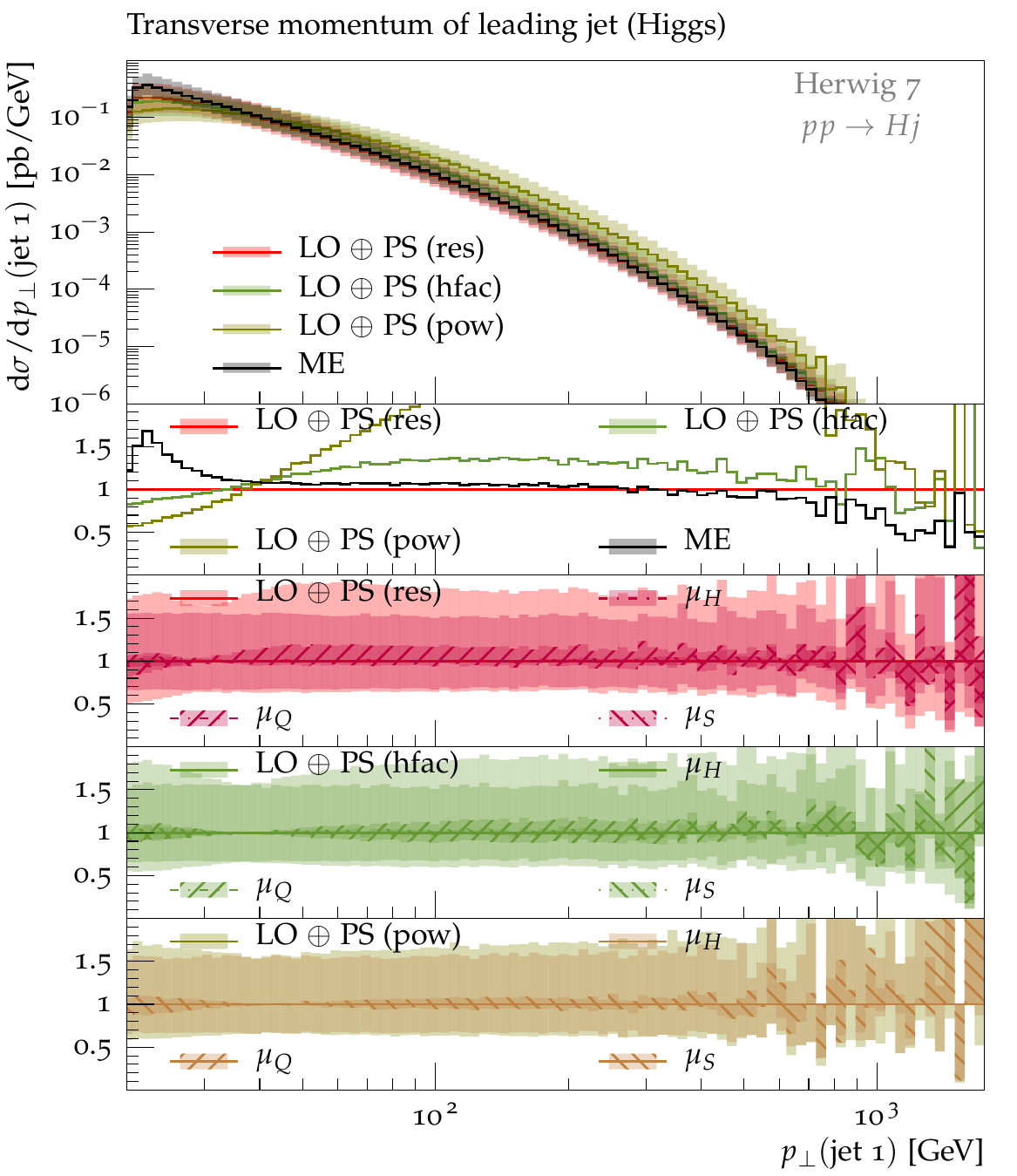}
\caption{Transverse momentum of the leading jet for Higgs plus one
  (inclusive) jet as computed by the QTilde shower for
  \texttt{resummation} (red), \texttt{hfact} (lime) and \texttt{power}
  (brown) profile compared to the ME (black) prediction. Top ratio
  plot: same as before. Other ratio plots: \texttt{resummation},
  \texttt{hfact} respectively \texttt{power} profile with full error band
  vs. variation of only either $\mu_H$, $\mu_Q$ or $\mu_S$.}
\label{fig:H_pT_def}
\end{figure}

\begin{figure}[t!]
\centering
\includegraphics[width=0.4\textwidth]{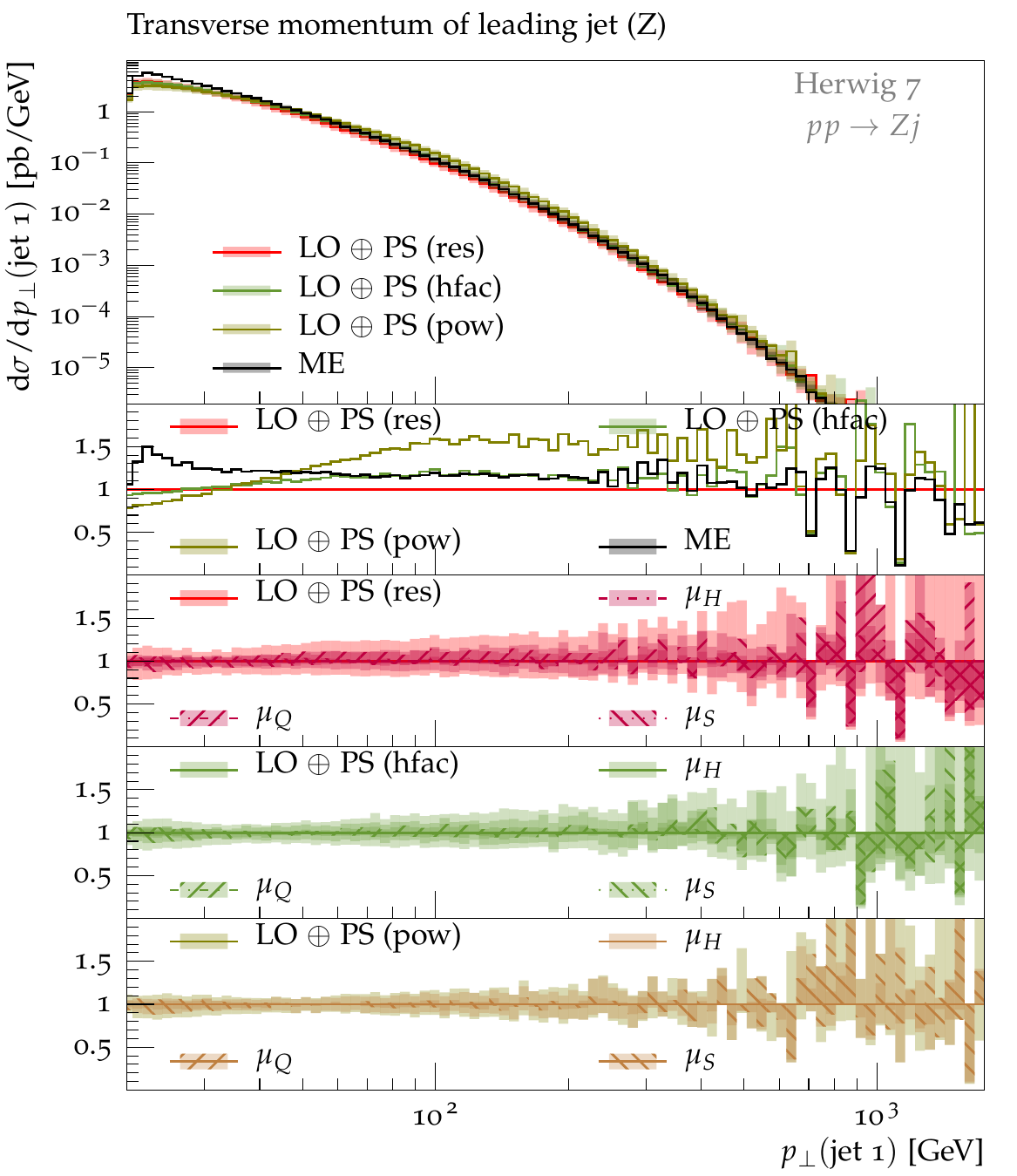}
\caption{Transverse momentum of the leading jet for Z plus one
  (inclusive) jet as computed by the QTilde shower for
  \texttt{resummation} (red), \texttt{hfact} (lime) and \texttt{power}
  (brown) profile compared to the ME (black) prediction. Top ratio
  plot: same as before. Other ratio plots: \texttt{resummation},
  \texttt{hfact} respectively \texttt{power} profile with full error band
  vs. variation of only either $\mu_H$, $\mu_Q$ or $\mu_S$.}
\label{fig:Z_pT_def}
\end{figure}

\begin{figure}[t!]
\centering
\includegraphics[width=0.4\textwidth]{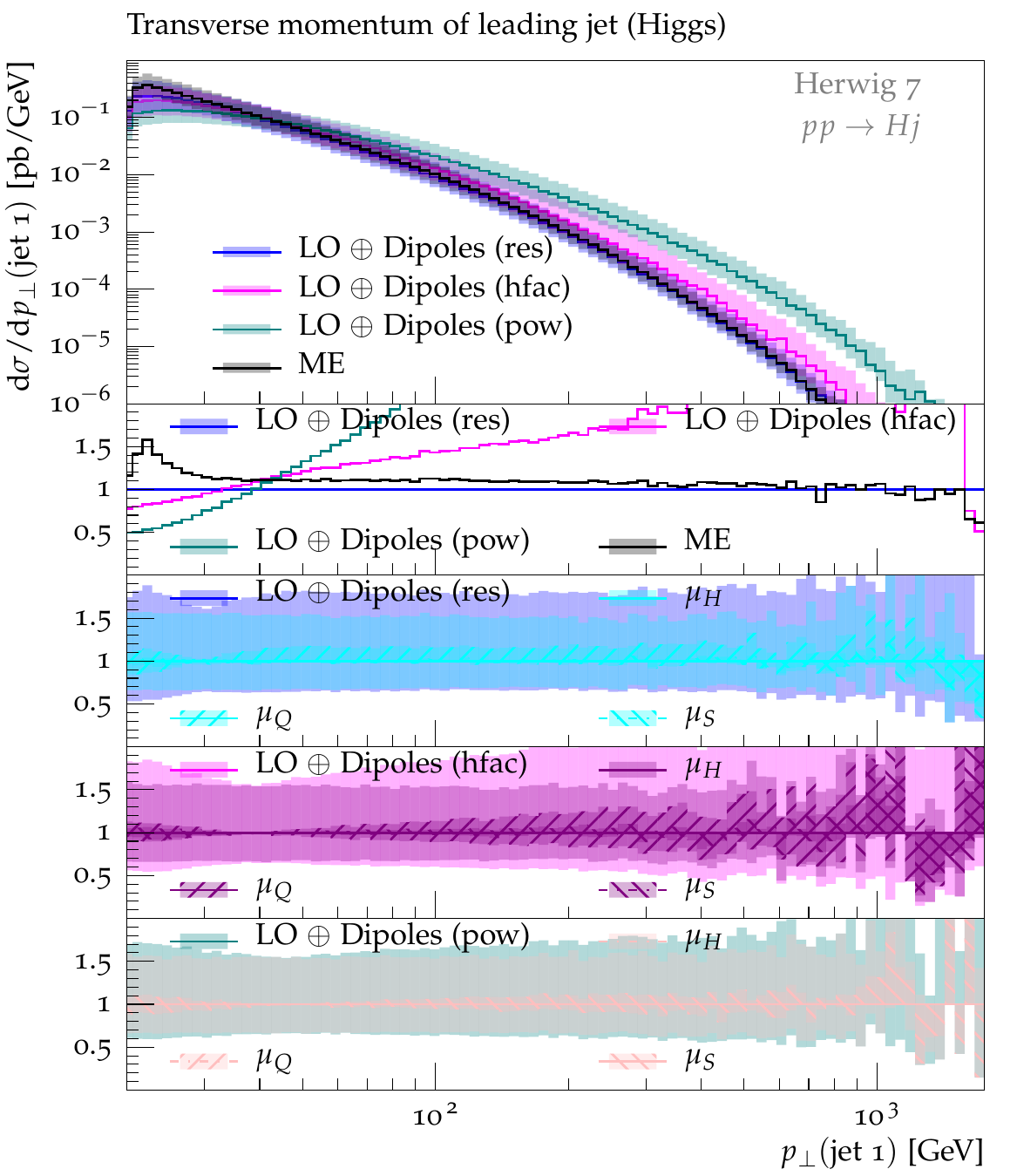}
\caption{Transverse momentum of the leading jet for Higgs plus one
  (inclusive) jet as computed by the Dipole shower for
  \texttt{resummation} (red), \texttt{hfact} (lime) and \texttt{power}
  (brown) profile compared to the ME (black) prediction. Top ratio
  plot: same as before. Other ratio plots: \texttt{resummation},
  \texttt{hfact} respectively \texttt{power} profile with full error band
  vs. variation of only either $\mu_H$, $\mu_Q$ or $\mu_S$.}
\label{fig:H_pT_dip}
\end{figure}

\begin{figure}[t!]
\centering
\includegraphics[width=0.4\textwidth]{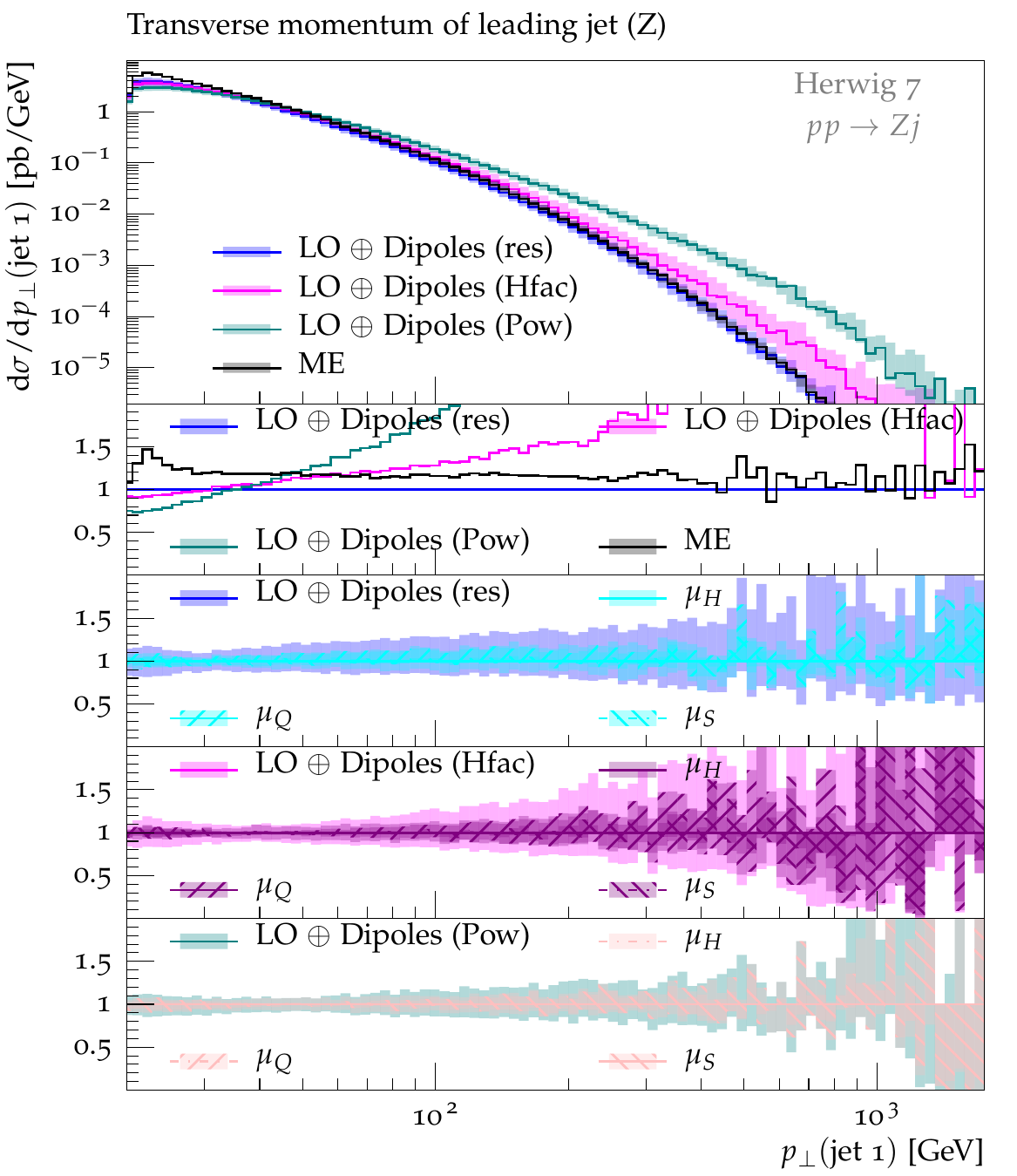}
\caption{Transverse momentum of the leading jet for Z plus one
  (inclusive) jet as computed by the Dipole shower for
  \texttt{resummation} (red), \texttt{hfact} (lime) and \texttt{power}
  (brown) profile compared to the ME (black) prediction. Top ratio
  plot: same as before. Other ratio plots: \texttt{resummation},
  \texttt{hfact} respectively \texttt{power} profile with full error band
  vs. variation of only either $\mu_H$, $\mu_Q$ or $\mu_S$.}
\label{fig:Z_pT_dip}
\end{figure}

Similarly to the rapidity distributions, we expect the $p_\perp$-spectra of
the leading jet to be predicted mainly by the hard matrix element,
according to the consistency conditions discussed in
Sec.~\ref{subsec:consistency}. In Fig.~\ref{fig:H_pT_def}
and~\ref{fig:Z_pT_def} (H and Z production, respectively) we show the results
for the QTilde shower, while Fig.~\ref{fig:H_pT_dip}
and~\ref{fig:Z_pT_dip} contain our findings for the Dipole shower. We
again find that the uncertainties are dominated by the variation of
the hard scale. For both showers the \texttt{resummation} profile is
consistent with the hard matrix element prediction, except for jets
close to the threshold where cut migration effects are being
probed\footnote{Cut migration for jetty processes should actually be
  considered another source of uncertainty beyond the ones discussed
  here; however, we do not address these in detail but chose to
  use equal generation and analysis cuts to highlight these
  effects.}. For the \texttt{hfact} profile with the QTilde shower we
find a spectrum compatible with the one anticipated by the matrix
element; for the dipole shower, a significantly harder spectrum is
obtained. A similar, but even more dramatic picture emerges for the
\texttt{power} shower setting.  The spread of predictions for the
QTilde shower is smaller than the spread for the dipole shower, owing
to the intrinsic limitations of the phase-space volume available to
angular-ordered emissions as already pointed out in the previous
sections. The combinations QTilde plus \texttt{power}, and
Dipole plus \texttt{hfact} or \texttt{power} contradict the
criterion of controllable showering, which in this case is expected to
not significantly alter the jet $p_\perp$ spectrum. Combined with the
empirical findings of Sec.~\ref{sec:clean}, we will therefore not
consider the \texttt{power} shower profile choice any further.
\begin{figure}[t!]
\centering
\includegraphics[width=0.4\textwidth]{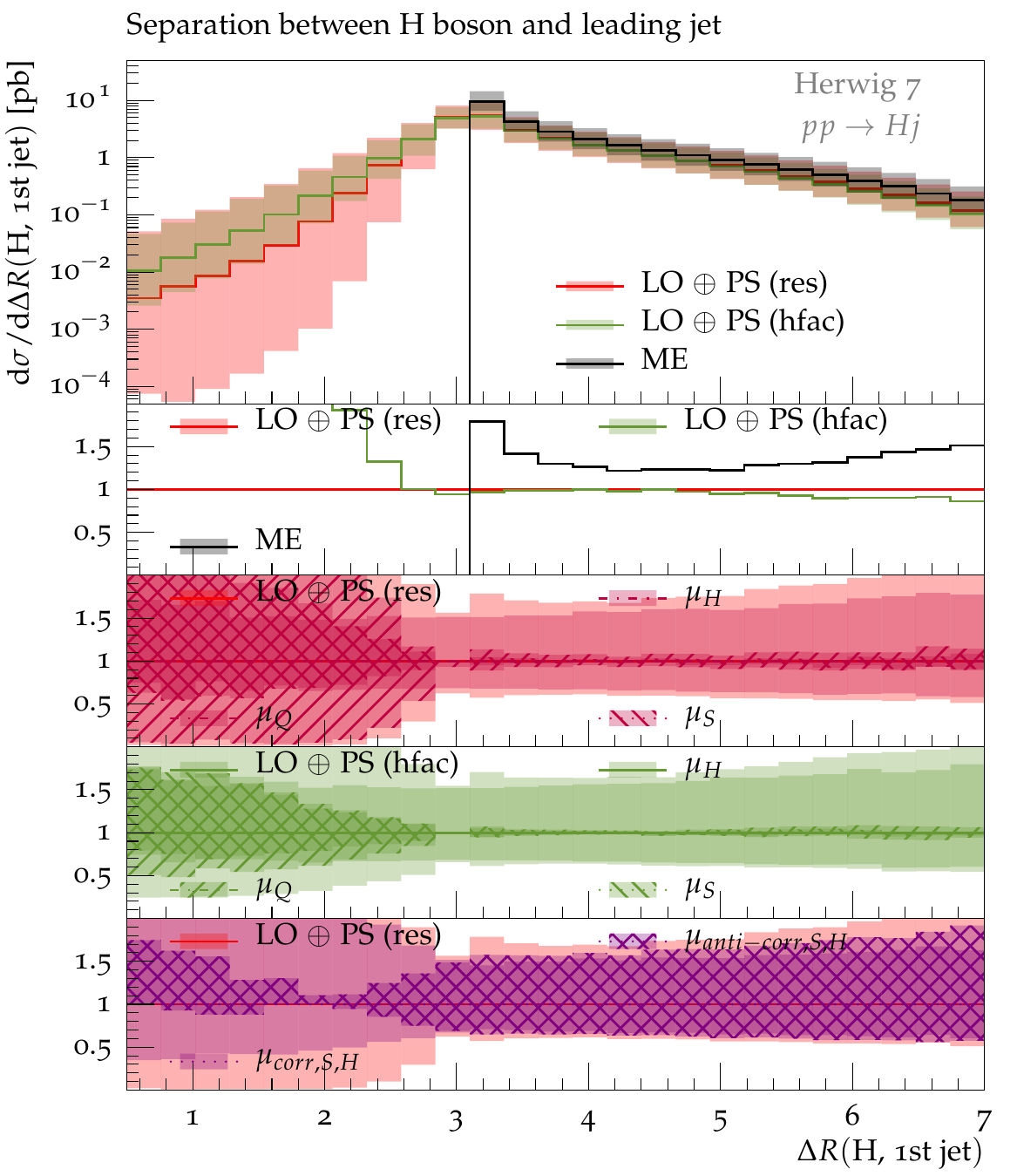}
\caption{Separation between the Higgs and the leading jet for Higgs
  plus one (inclusive) jet as computed by the QTilde shower for
  \texttt{resummation} (red) and \texttt{hfact} (lime) profile compared to
  the ME (black) prediction. Top ratio plot: same as before. Second
  and third ratio plots: \texttt{resummation} respectively \texttt{hfact}
  profile with full error band vs. variation of only either $\mu_H$,
  $\mu_Q$ or $\mu_S$. Last ratio plot: \texttt{resummation} profile with
  full error band vs a subset where $\mu_H$ and $\mu_S$ are varied in
  correlated (dark purple) respectively anti-correlated (hatched)
  manner, while $\mu_Q$ is held fixed.}
\label{fig:H_dR_def}
\end{figure}
\begin{figure}[t!]
\centering
\includegraphics[width=0.4\textwidth]{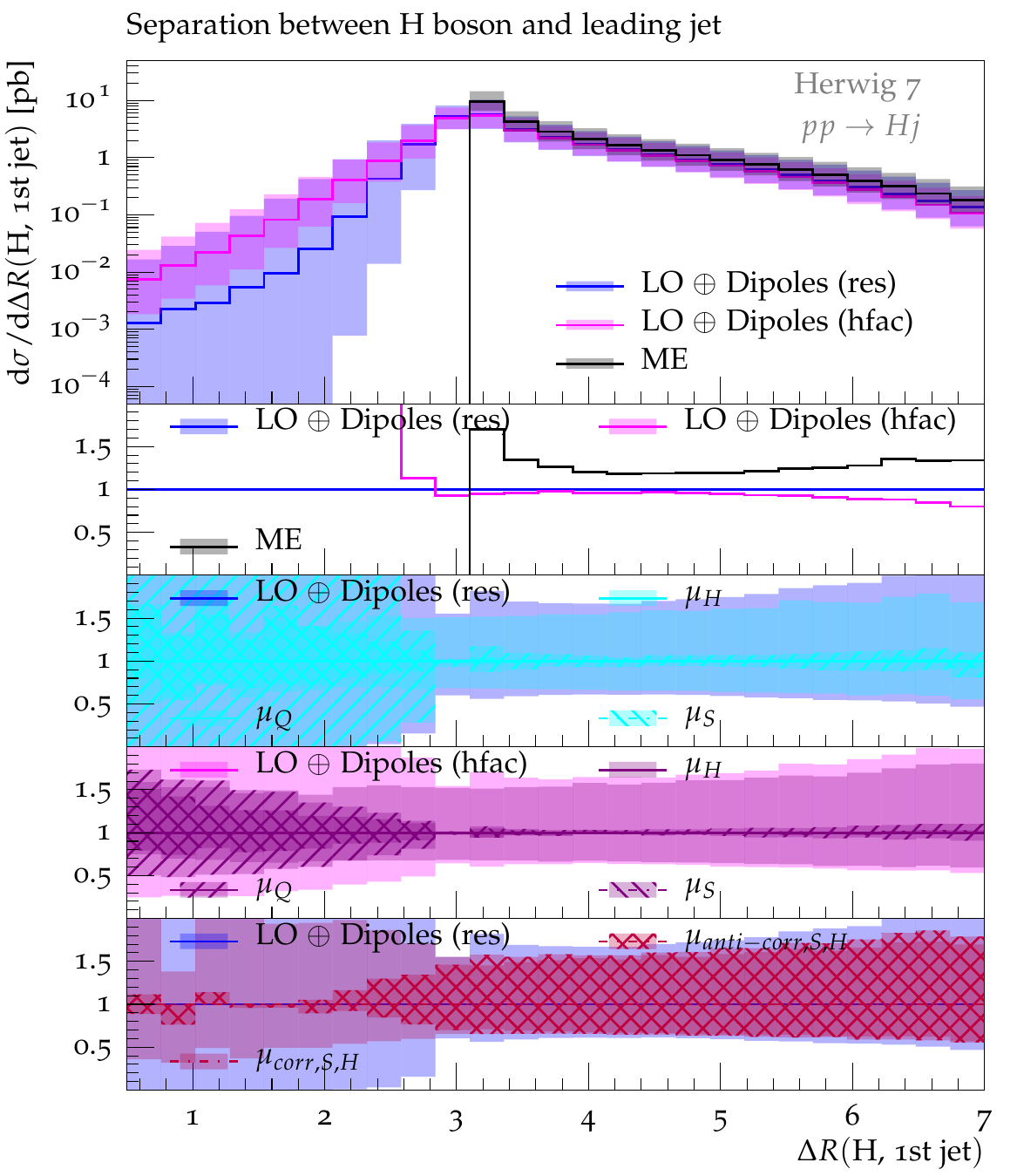}
\caption{Separation between the Higgs and the leading jet for Higgs
  plus one (inclusive) jet as computed by the Dipole shower for
  \texttt{resummation} (red) and \texttt{hfact} (lime) profile compared to
  the ME (black) prediction. Top ratio plot: same as before. Second
  and third ratio plots: \texttt{resummation} respectively \texttt{hfact}
  profile with full error band vs. variation of only either $\mu_H$,
  $\mu_Q$ or $\mu_S$. Last ratio plot: \texttt{resummation} profile with
  full error band vs. a subset where $\mu_H$ and $\mu_S$ are varied in
  a correlated (dark purple) respectively anti-correlated (hatched)
  manner, while $\mu_Q$ is held fixed.}
\label{fig:H_dR_dip}
\end{figure}
\begin{figure}[t!]
\centering
\includegraphics[width=0.4\textwidth]{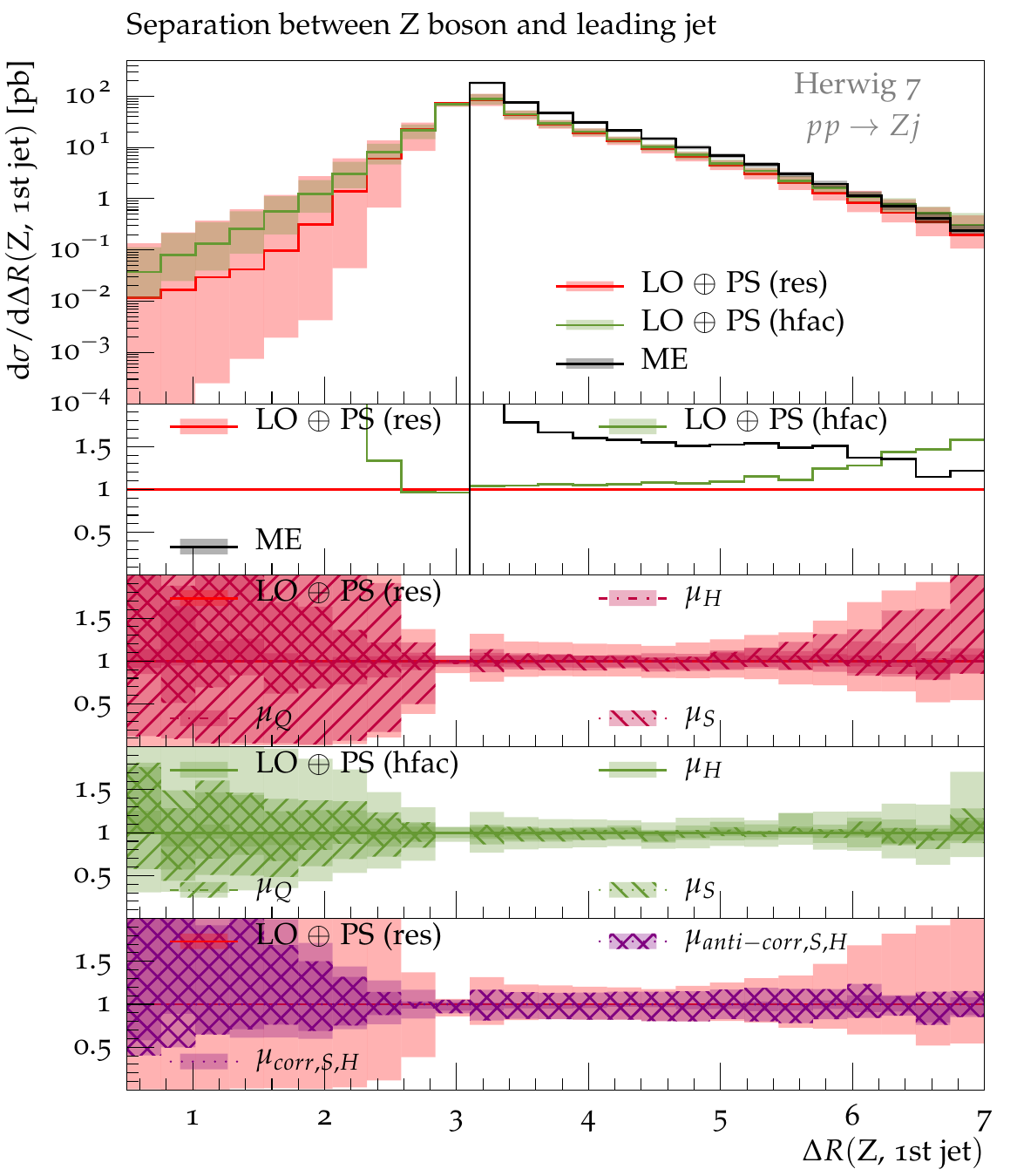}
\caption{Separation between the Z and the leading jet for Z plus one
  (inclusive) jet as computed by the QTilde shower for
  \texttt{resummation} (red) and \texttt{hfact} (lime) profile compared to
  the ME (black) prediction. Top ratio plot: same as before. Second
  and third ratio plots: \texttt{resummation} respectively \texttt{hfact}
  profile with full error band vs. variation of only either $\mu_H$,
  $\mu_Q$ or $\mu_S$. Last ratio plot: \texttt{resummation} profile with
  full error band vs a subset where $\mu_H$ and $\mu_S$ are varied in
  a correlated (dark purple) respectively anti-correlated (hatched)
  manner, while $\mu_Q$ is held fixed.}
\label{fig:Z_dR_def}
\end{figure}
\begin{figure}[t!]
\centering
\includegraphics[width=0.4\textwidth]{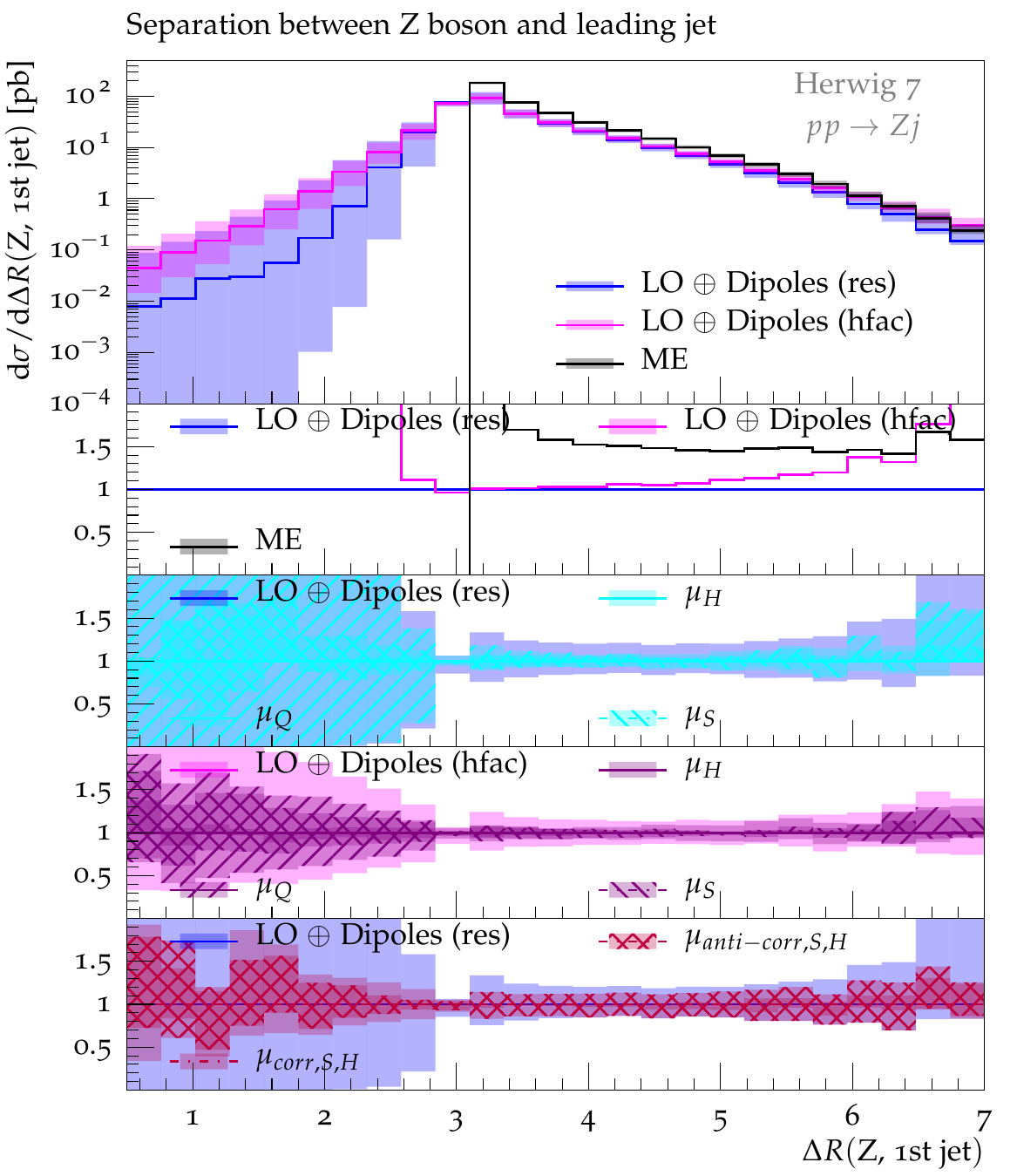}
\caption{Separation between the Z and the leading jet for Z plus one
  (inclusive) jet as computed by the Dipole shower for
  \texttt{resummation} (red) and \texttt{hfact} (lime) profile compared to
  the ME (black) prediction. Top ratio plot: same as before. Second
  and third ratio plots: \texttt{resummation} respectively \texttt{hfact}
  profile with full error band vs. variation of only either $\mu_H$,
  $\mu_Q$ or $\mu_S$. Last ratio plot: \texttt{resummation} profile with
  full error band vs a subset where $\mu_H$ and $\mu_S$ are varied in
  a correlated (dark purple) respectively anti-correlated (hatched)
  manner, while $\mu_Q$ is held fixed.}
\label{fig:Z_dR_dip}
\end{figure}

Turning to more exclusive observables\footnote{We remind the reader
  that `exclusive' here means: potentially probing more and more
  shower emissions on top of the hard process.}, we consider the
angular separation between the boson and the leading jet $\Delta
R_{(H/Z)j}$, which probes both matrix element and shower dominated
regions in a continuous observable: Matrix element emissions in this
case can only populate the phase-space region $\Delta R_{(H/Z)j} \ge
\pi$. The region below is solely filled by the parton shower,
typically operating at the boundary of validity of the underlying
approximation as this phase space requires the shower to produce a
hard, large-angle emission. Within the definition of controllable and
consistent uncertainties, we therefore expect large uncertainties for
$\Delta R_{(H/Z)j} \le \pi$, while the shower should reproduce the
matrix element dynamics above.  Results for the QTilde shower are
shown in Fig.~\ref{fig:H_dR_def} and~\ref{fig:Z_dR_def} (H and
$Z$ production, respectively) and for the Dipole shower in
Fig.~\ref{fig:H_dR_dip} and~\ref{fig:Z_dR_dip}. We place particular
emphasis on the comparison of the \texttt{resummation} and
\texttt{hfact} profiles. For all processes/showers we find that
\texttt{hfact} predicts a small uncertainty band and produces slightly
more hard jets; the latter can be attributed to the available phase
space, while the former can be obtained by analysing
Eq.~\ref{eqns:logstructure}, stressing the fact that the region in
which the derivative of the profile is varying significantly extends
over a larger region than for the other profiles, though with less
overall variation implied. Contrary, and matching the expectations
motivated by the logarithmic structure, the uncertainty for the
\texttt{resummation} profile in the small $\Delta R_{(H/Z)j}$ region
is large and driven by all scale variations together. In addition, in
the bottom ratio plot of Figs.~\ref{fig:H_dR_def}, \ref{fig:Z_dR_def},
\ref{fig:H_dR_dip} and \ref{fig:Z_dR_dip} we show a subset of scale
variations for the \texttt{resummation} profile choice, varying the
hard and shower scales in a correlated and anti-correlated setting, at
a fixed $\mu_Q$. This breakdown shows how different subsets of
variations constitute the full uncertainty band. Besides the simple
domination of the uncertainty by one variation, other regions of phase
space show that the uncertainty is strongly underestimated by
considering the variations separately. We therefore argue that only
the full, combined, scale variation produces a reliable error band.
\begin{figure}[t!]
\centering
\includegraphics[width=0.4\textwidth]{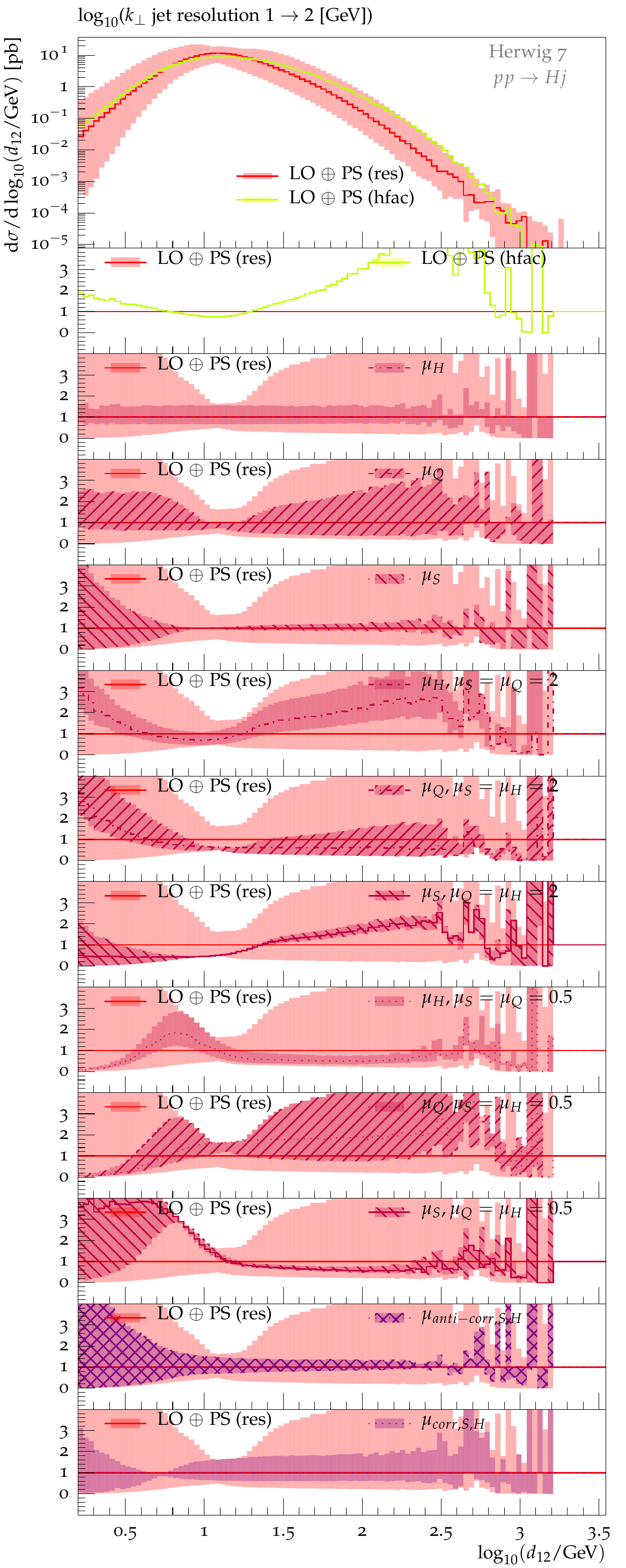}
\caption{$k_\perp$--splitting scale for the transition from the one to the
  two jet configuration in Higgs plus one (inclusive) jet as computed
  with the QTilde shower. The base line in each plot is the
  \texttt{resummation} profile choice. The main plot and the first ratio
  plot show a comparison to the \texttt{hfact} profile choice. The
  subsequent ratio plots compare the full error band to certain
  subsets of scale variation choices, namely: variation of $\mu_H$,
  $\mu_Q$, $\mu_S$, $\mu_H$ with $\xi_Q$ and $\xi_S$ fixed to 2,
  $\mu_Q$ with $\xi_H$ and $\xi_S$ fixed to 2, $\mu_S$ with $\xi_Q$
  and $\xi_H$ fixed to 2, $\mu_H$ with $\xi_Q$ and $\xi_S$ fixed to
  0.5, $\mu_Q$ with $\xi_H$ and $\xi_S$ fixed to 0.5, $\mu_S$ with
  $\xi_Q$ and $\xi_H$ fixed to 0.5, $\mu_S$ and $\mu_H$ in a
  correlated manner and $\mu_S$ and $\mu_H$ in an anti-correlated
  manner.}
\label{fig:H_d_12_def}
\end{figure}
As another probe of the interaction of shower emissions with the hardest jet, we
consider $k_\perp$-splitting scales, particularly the one in which an event with two
jets would turn into an event with one jet as the jet $p_\perp$ threshold passes
through the scale obtained. These observables have also been proven to
be accessible to analytic considerations~\cite{Gerwick:2014koa}, for which
comparisons to full parton showers are highly desirable though are beyond the
scope of this paper. In Fig.~\ref{fig:H_d_12_def} we show our results for
the QTilde shower for Higgs production\footnote{The results for Z plus one jet and the Dipole
  shower yield similar observations and conclusions.}. Once again we compare the
\texttt{resummation} profile choice with the \texttt{hfact} profile. It is
noteworthy that the \texttt{hfact} profile introduces a strong change in the
shape of the Sudakov peak, on top of the harder spectrum already observed for
the first jet; besides the tail effects we are therefore concerned that
profile scale choices along these lines may significantly impact the
resummation properties of the parton shower, as may already be expected from
the arguments presented in Sec.~\ref{sec:scales}. We therefore
conclude that, even with intrinsically restricted phase space, the
\texttt{hfact} profile does not provide controllable uncertainties and
will not be taken further into account in this study. We also use
Fig.~\ref{fig:H_d_12_def} to perform an comprehensive breakdown of the different
variation directions in the `cube' of possible variations, showing that no
individual variation actually covers the full dynamics present. For LO
plus PS simulations, we therefore argue that the full band is taken into
consideration and improvements in the context of matching and merging will be
subject to future work.
\begin{figure}[t!]
\centering
\includegraphics[width=0.4\textwidth]{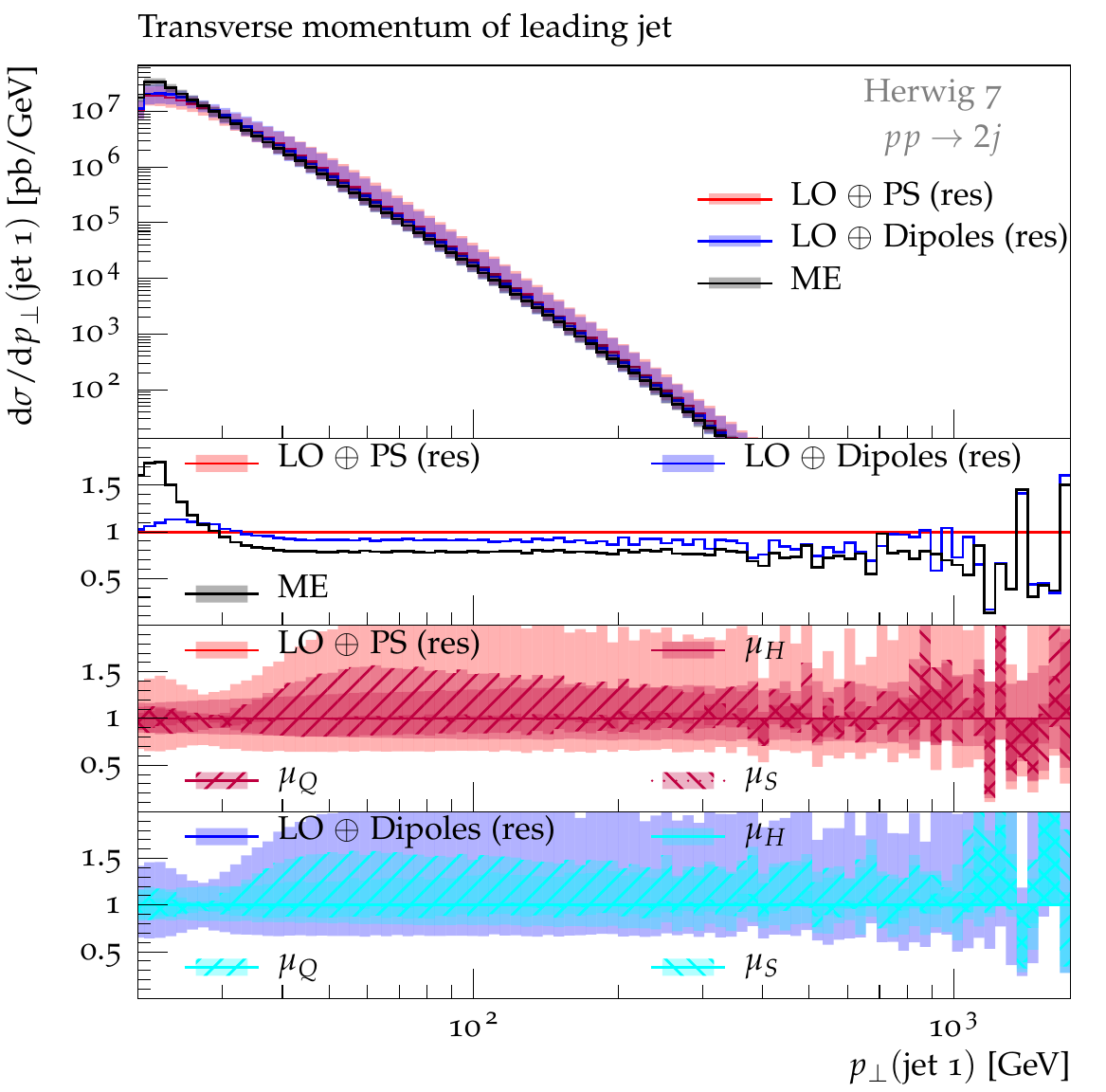}
\caption{Transverse momentum of the first jet in (inclusive) di-jet
  production as computed by the QTilde (red) and Dipole (blue) shower
  with the \texttt{resummation} profile compared to the ME (black)
  prediction. Top ratio plot: same as before. Subsequent ratio plots:
  QTilde (second) and Dipole (third) with full error band
  vs. variation of only either $\mu_H$, $\mu_Q$ or $\mu_S$.}
\label{fig:j_pt1}
\end{figure}
\begin{figure}[t!]
\centering
\includegraphics[width=0.4\textwidth]{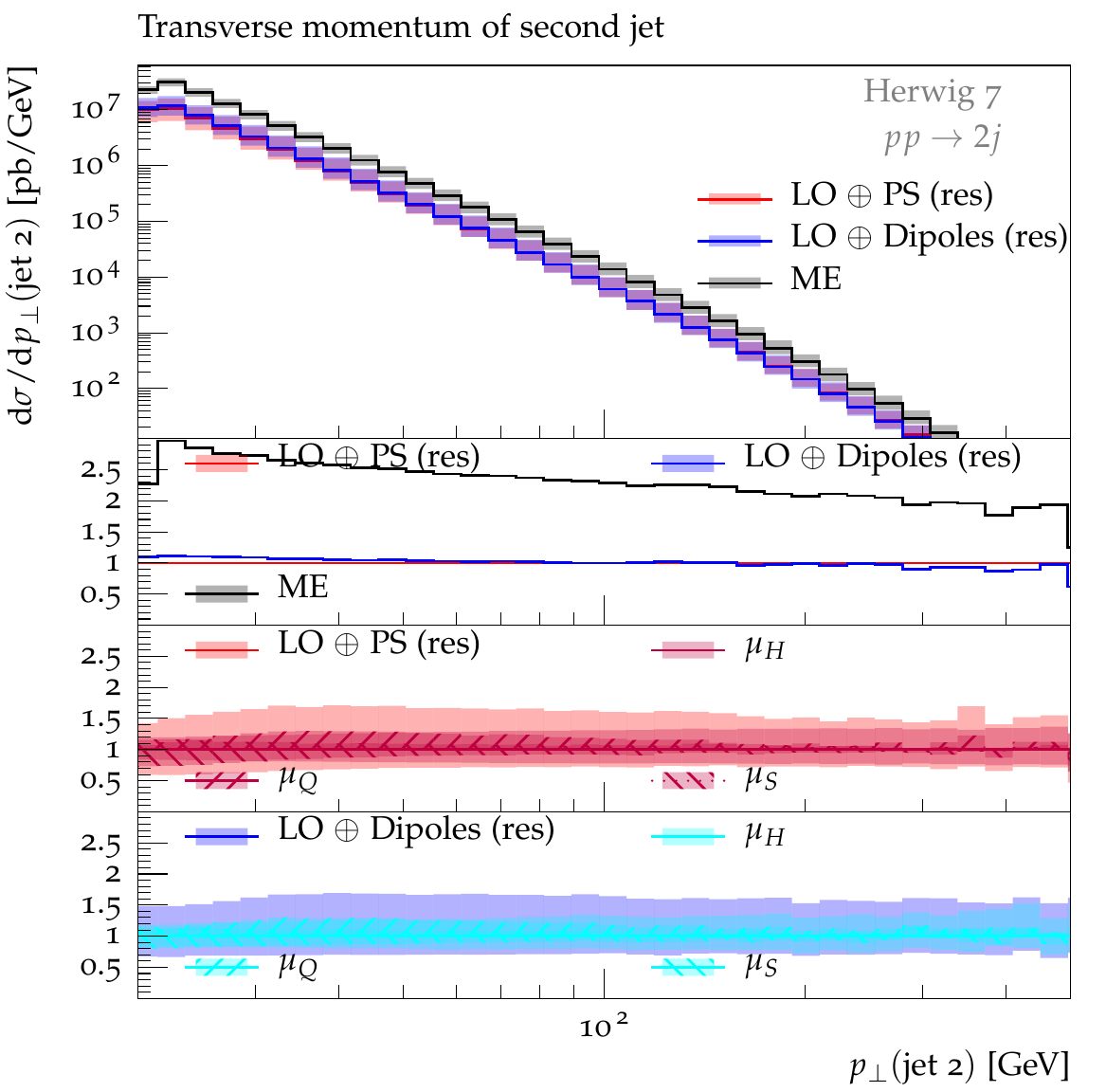}
\caption{Transverse momentum of the second jet in (inclusive) di-jet
  production as computed by the QTilde (red) and Dipole (blue) shower
  with the \texttt{resummation} profile compared to the ME (black)
  prediction. Top ratio plot: same as before. Subsequent ratio plots:
  QTilde (second) and Dipole (third) with full error band
  vs. variation of only either $\mu_H$, $\mu_Q$ or $\mu_S$.}
\label{fig:j_pt2}
\end{figure}
\begin{figure}[t!]
\centering
\includegraphics[width=0.4\textwidth]{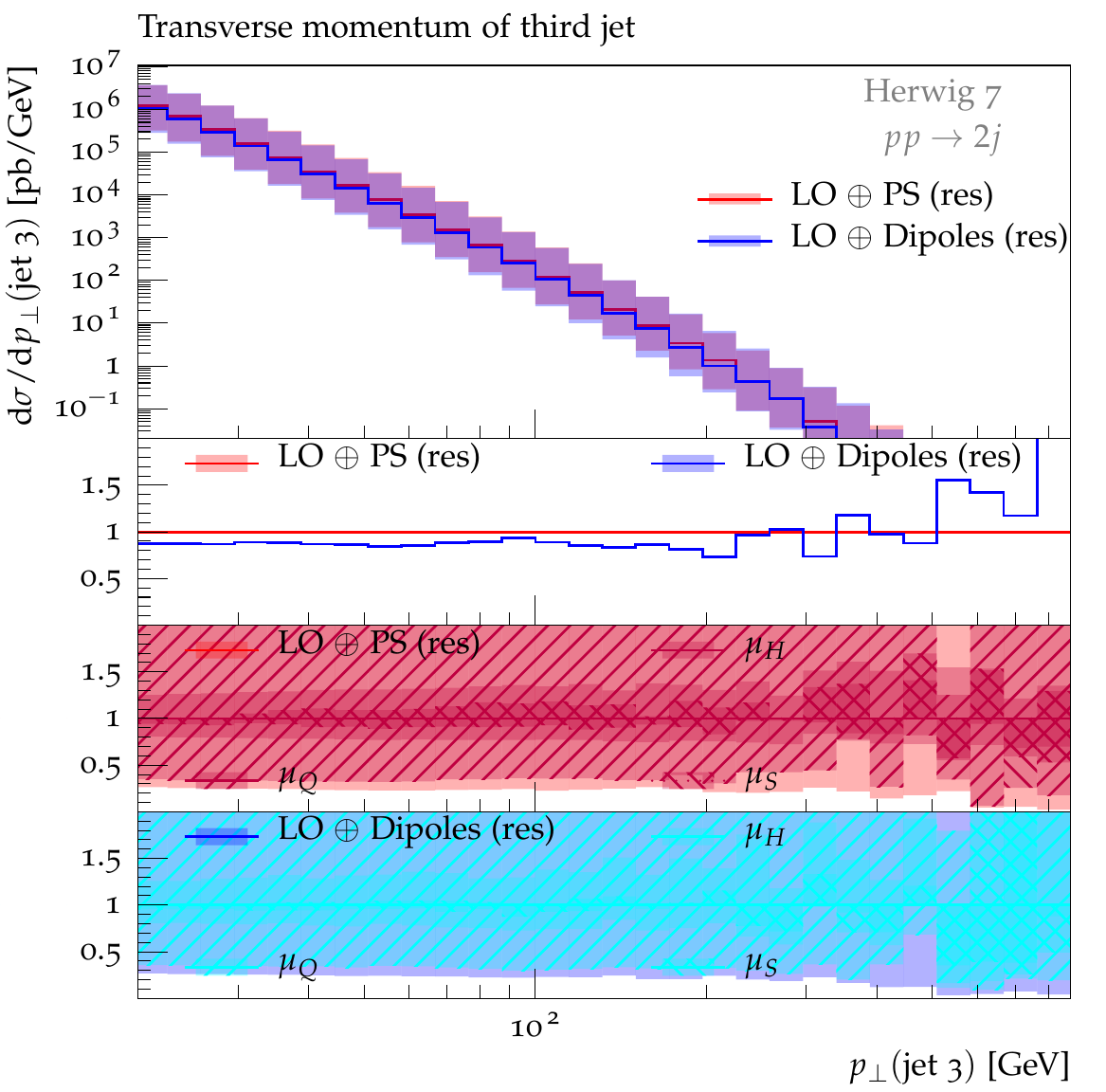}
\caption{Transverse momentum of the third jet (inclusive) di-jet
  production as computed by the QTilde (red) and Dipole (blue) shower
  with the \texttt{resummation} profile compared to the ME (black)
  prediction. Top ratio plot: same as before. Other ratio plots:
  QTilde (second) and Dipole (third) with full error band
  vs. variation of only either $\mu_H$, $\mu_Q$ or $\mu_S$.}
\label{fig:j_pt3}
\end{figure}

\begin{figure}[t!]
\centering
\includegraphics[width=0.4\textwidth]{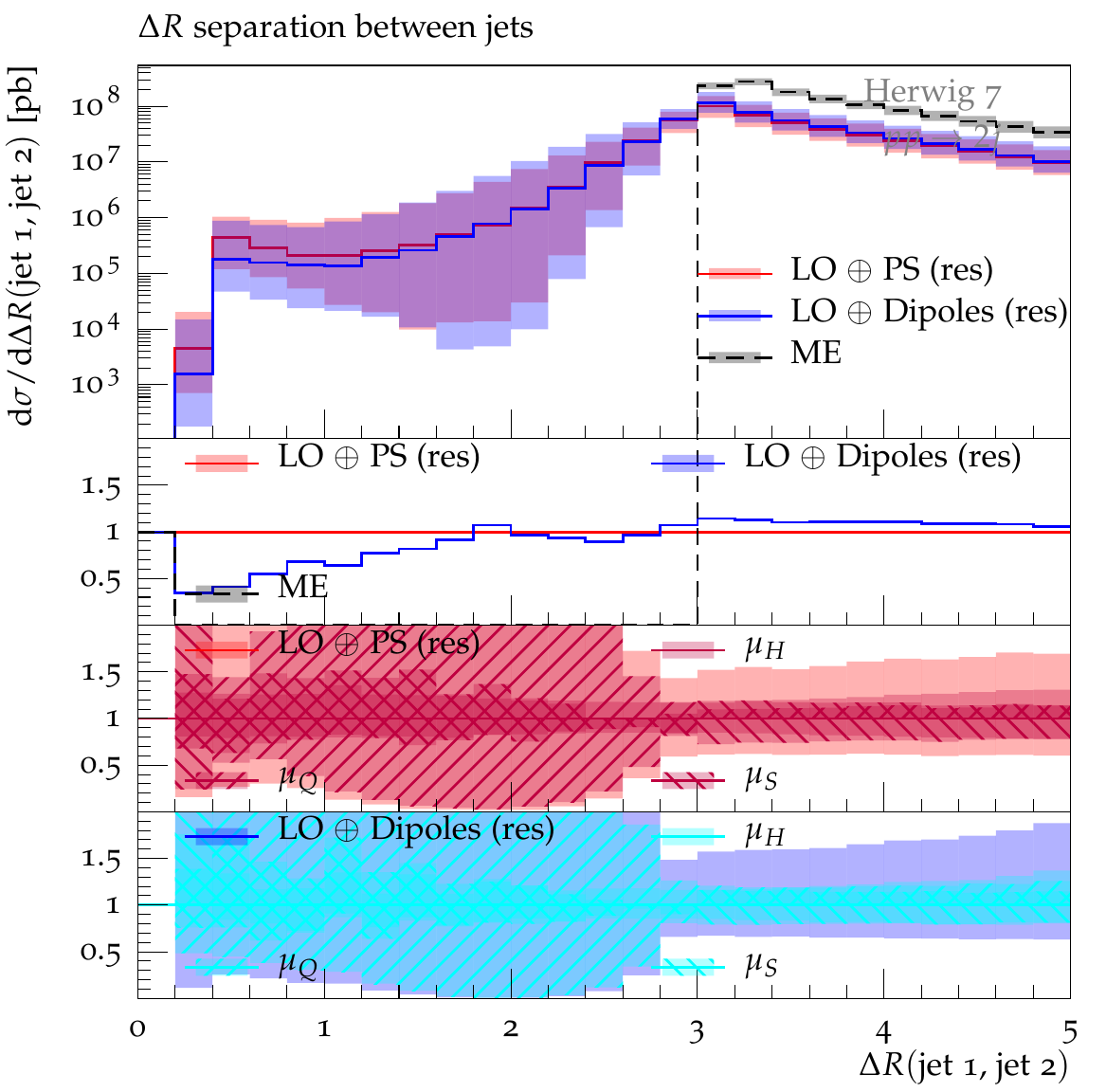}
\caption{Separation between the leading and second jet for (inclusive)
  di-jet as computed by the QTilde (red) and the Dipole (blue) shower
  together with the \texttt{resummation} profile. Top ratio plot: same
  as before. Second and third ratio plots: full error band
  vs. variation of only either $\mu_H$, $\mu_Q$ or $\mu_S$.}
\label{fig:j_dR_12}
\end{figure}

We have so far considered processes with a colourless, massive object that
dominates the scale hierarchy at hand, and, even in the presence of an
additional jet, makes the dynamics rather insensitive to additional radiation
(as far as this radiation is confined to reasonable phase-space regions as
identified above). A process where this is clearly not the case is pure jet
production in hadron collisions, which also probes different colour structures
that have not been encountered in the hard processes considered thus far.
Owing to the back-to-back configuration at lowest order, we expect
considerable parton-shower effects in comparison to the hard matrix element
for a number of observables and expect to make a more detailed comparison to
fixed order only once NLO improvement has been incorporated. Nevertheless, we
can still test as to what extent the shower variations match up to
expectations in signalling regions where the prediction should generally be
considered unreliable. We also test, once more, if the two showers are
comparable within their uncertainties. Following the previous arguments, we
only consider the \texttt{resummation} profile, with a hard scale again given
by the jet $p_\perp$. Sample results comparing to the hard matrix element are
shown in Figs.~\ref{fig:j_pt1}, \ref{fig:j_pt2} and \ref{fig:j_dR_12}, which 
show that the two showers preform in a very similar way both in their
central predictions and variations; they also show that {\it qualitatively} we
find a behaviour similar to the singlet plus jet benchmarks as if we had
replaced the hard, colourless, object with a jet as hard probe. {\it
  Quantitatively}, however, we observe significant changes in rates for the
second jet, which need to be confronted with the impact of cut migration as
well as the impact of higher order corrections. We also note that choosing the
hard veto scale in this setting has a significant impact on showered results.

\begin{figure}[t!]
\centering
\includegraphics[width=0.4\textwidth]{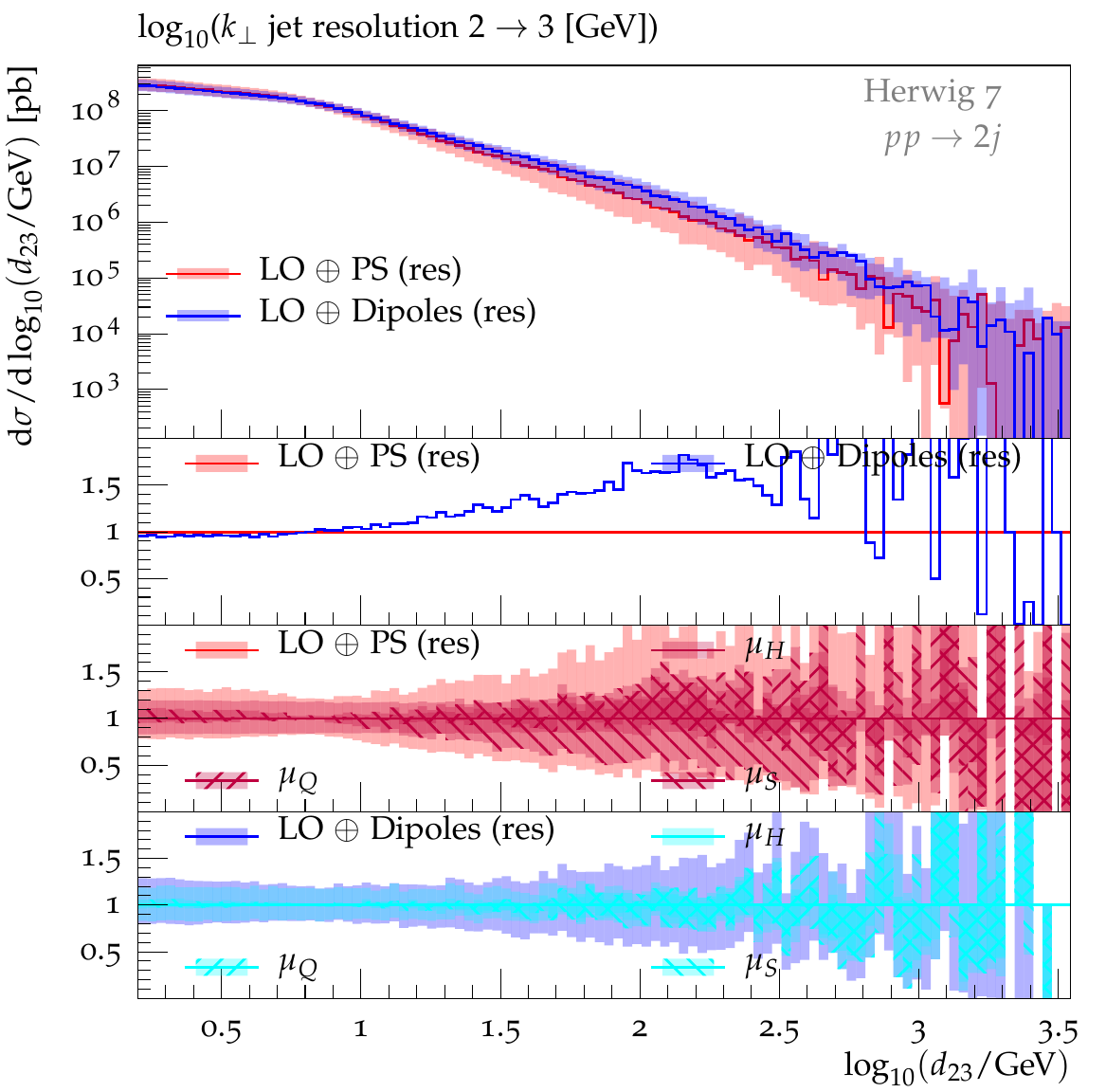}
\caption{$k_\perp$--splitting scale for the transition from the two to the
  three jet configuration in (inclusive) di-jet production as computed
  with the QTilde (red) and the Dipole (blue) shower together with the
  \texttt{resummation} profile. The base line in the first plot is the
  QTilde shower. The subsequent ratio plots compare the full error
  band to certain subsets of scale variation choices, namely:
  variation of $\mu_H$, $\mu_Q$, $\mu_S$. The second ratio plot
  presents the QTilde shower, the last one the Dipole shower.}
\label{fig:j_d23}
\end{figure}

With the transverse momentum of the third jet and the $2\to 3$ resolution
shown in Figs.~\ref{fig:j_pt3} and \ref{fig:j_d23} we consider purely shower
driven quantities; both of these nicely reveal that the two showers, together
with the \texttt{resummation} profile, are perfectly compatible with each other,
exhibiting the same resummation accuracy.

\section{Conclusions and Outlook}
\label{sec:outlook}

We have performed a comprehensive and detailed study of the sources of
uncertainty in parton showers, utilising the two parton-shower algorithms
available in \hw. We have investigated different choices of profile scales to
approach the boundary of hard emissions, as these are highly relevant to
effects that appear in the context of NLO plus PS matching. We have
systematically categorised the sources of uncertainty and outlined their
interplay with other simulation components, putting this study into context of a
bigger work programme to eventually establish uncertainties for event
generators in total.

Focussing on the perturbative, parton-shower part, of the simulation, we have
deliberately chosen LO plus PS calculations to establish a baseline of
controllable and consistent variations that will allow us to subsequently
identify improvements and reduction in these uncertainties as higher order
corrections are included. We have found that profile scale choices are very
constrained when applying consistency conditions on both central predictions
(which should not alter input distributions of the hard process) and
uncertainties (with large uncertainties to be expected in unreliable regions
or regions dominated by hadronisation corrections), as well as stable results
in the Sudakov region. Particularly the \texttt{hfact} and \texttt{power}
shower configurations do not admit results compatible with these criteria.
Utilising a \texttt{resummation} profile, which is very close to the
\texttt{theta} cutoff for hard emissions as implemented in previous
algorithms, we find that the angular-ordered and dipole-based shower
algorithms are compatible with each other, both in central predictions and
uncertainty claims, despite their very different nature. 

\section*{Acknowledgments}

We are grateful to the other members of the Herwig collaboration for
encouragement and helpful discussions; in particular we would like to thank
Stefan Gieseke, Peter Richardson and Mike Seymour for a careful review of the
manuscript.  We also acknowledge fruitful exchange with Mrinal Dasgupta and
Frank Tackmann.

The work of JB and PS has been supported by the European Union as part of the
FP7 Marie Curie Initial Training Network MCnetITN (PITN-GA-2012-315877). GN
acknowledges a short term student visit funded by MCnetITN. SP acknowledges
support by a FP7 Marie Curie Intra European Fellowship under Grant Agreement
PIEF-GA-2013-628739. We are also grateful to the Cloud Computing for Science
and Economy project (CC1) at IFJ PAN (POIG 02.03.03-00-033/09-04) in Cracow
and the U.K. GridPP project whose resources were used to carry out some of the
numerical calculations for this project.

Thanks also to Mariusz Witek and Mi\l{}osz Zdyba\l{} for their help with 
CC1 and Oliver Smith for his help with grid computing.

\bibliography{shower-uncertainties}

\end{document}